\documentclass[11pt,a4paper]{article} 
\pdfoutput=1

\usepackage{jheppub}

\usepackage{amsmath,amssymb}
\usepackage{multirow}
\usepackage{graphicx}
\usepackage{subfigure}
\usepackage{tikz}
\usepackage{color}

\DeclareMathOperator{\diag}{diag}
\DeclareMathOperator{\imag}{Im}
\DeclareMathOperator{\tr}{Tr}

\usetikzlibrary{shapes,decorations.markings}
\tikzset{
  gauge/.style={decorate, decoration={snake}},
  scalar/.style={dashed},
  fermion/.style={postaction={decorate},
    decoration={markings,mark=at position .55 with {\arrow{>}}}},
  gluon/.style={decorate, decoration={coil,amplitude=4pt, segment length=5pt}} 
}

\title{The minimal adjoint-$\boldmath{\mathbf{SU(5)}\times\mathbf{Z_4}}$ GUT model}

\author[a]{D.~Emmanuel-Costa,}
\author[a]{C.~Sim\~oes,}
\author[b]{and M.~T\'ortola}

\emailAdd{david.costa@ist.utl.pt}
\emailAdd{csimoes@cftp.ist.utl.pt}
\emailAdd{mariam@ific.uv.es}

\affiliation[a]{
Departamento de F\'{\i}sica and Centro de F\'{\i}sica Te\'orica de Part\'{\i}ıculas (CFTP)\\
Instituto Superior T\'ecnico, Universidade T\'ecnica de Lisboa\\
Av. Rovisco Pais, P-1049-001 Lisboa, Portugal.
}
\affiliation[b]{AHEP Group, Instituto de F\'{\i}sica Corpuscular - 
  C.S.I.C./Universitat de Val\`encia,\\
  Edificio Institutos de Paterna, Apt 22085, E-46071 Valencia, Spain}

\abstract{ An extension of the adjoint $\mathsf{SU(5)}$ model with a
  flavour symmetry based on the $\mathsf{Z}_4$ group is
  investigated. The $\mathsf{Z}_4$ symmetry is introduced with the aim
  of leading the up- and down-quark mass matrices to the
  Nearest-Neighbour-Interaction form. As a consequence of the discrete
  symmetry embedded in the $\mathsf{SU(5)}$ gauge group, the charged
  lepton mass matrix also gets the same form. Within this model, light
  neutrinos get their masses through type-I, type-III and one-loop
  radiative seesaw mechanisms, implemented, respectively, via a
  singlet, a triplet and an octet from the adjoint fermionic
  $\mathsf{24}$ fields. It is demonstrated that the neutrino
  phenomenology forces the introduction of at least three
  $\mathsf{24}$ fermionic multiplets.  The symmetry
  $\mathsf{SU(5)}\times\mathsf{Z}_4$ allows only two viable zero
  textures for the effective neutrino mass matrix. It is showed that
  one texture is only compatible with normal hierarchy and the other
  with inverted hierarchy in the light neutrino mass
  spectrum. Finally, it is also demonstrated that $\mathsf{Z}_4$
  freezes out the possibility of proton decay through exchange of
  colour Higgs triplets at tree-level.  }

\preprint{CFTP/13-008, IFIC/13-10}

\arxivnumber{arXiv:1303.XXXX}

\begin{document}

\maketitle
\flushbottom


\section{Introduction}
\label{sec:intro}

Grand Unified Theories (GUT) are natural extensions of the Standard
Model (SM) and provide an appealing framework for the search of the
theory of flavour. Most GUT models try to unify the three gauge
couplings of SM in a unique coupling within a simple group. This is
sustained by the fact that the SM gauge couplings seem to unify at
high scale, $\Lambda\approx10^{15-17}$ GeV, when they evolve through
the renormalisation group equations. In such GUT constructions, not
only the SM gauge coupling unify, but also the SM fermions are tight
in larger multiplets opening the possibility for the implementation of
a flavour symmetry. Another important signature of most GUTs is the
prediction for proton decay~\cite{Nath:2006ut}, which has not yet been
observed and severely constrains these models.

The first GUT model was realisable within the $\mathsf{SU}(5)$ gauge
group~\cite{Georgi:1974sy} in 1974. This minimal model fits the
fifteen SM fermionic degrees of freedom in two unique
representations:~$\mathsf{5}^{\ast}$ and $\mathsf{10}\,$, per
generation. It is well established that this model is ruled out, since
it does not reproduce the correct mass ratios among the charged
leptons and down-type quarks and also the particle content does not
lead to an accurate gauge coupling unification. During the last
decades, many attempts have been proposed in the literature in order
to construct consistent GUT
models~\cite{Bajc:2002bv,Bajc:2002pg,EmmanuelCosta:2003pu,Perez:2007rm}
based on the $\mathsf{SU}(5)$ group. In particular, the mass mismatch
between the charged leptons and down-type quarks in the minimal
$\mathsf{SU}(5)$ can be easily corrected if one accepts higher
dimension operators in the model without enlarging the field
content~\cite{Bajc:2002bv,EmmanuelCosta:2003pu}. Alternatively, one
can build a non-renormalisable solution where the mass mismatch is
explained by adding an extra $\mathsf{45}$ Higgs
multiplet~\cite{Georgi:1979df,Langacker:1980js}.

Although GUT multiplets contain both quark and lepton fields this is
not enough to fully determine the properties of their observed masses
and mixings. Indeed, GUT models do not solve the ``flavour puzzle''
present in the SM, however the new GUT relations among quark and
lepton Yukawa matrices are an excellent starting point for building a
flavour symmetry. There have been many approaches to understand the
intricate ``flavour puzzle'' in the context of GUTs. An attractive
possibility is to assume the vanishing of some Yukawa interactions by
the requirement of a discrete symmetry, so that new ``texture zeroes''
appear in the Yukawa matrices~\cite{Branco:1987tv, Babu:2004tn,
  Grimus:2004hf, Low:2005yc, Ferreira:2010ir,
  Canales:2011ug}. Symmetries may predict new relations among fermion
masses and their mixings. Nevertheless, the opposite is not true in
general, because zeroes in the Yukawa matrices can also be obtained by
performing some set of transformations (weak basis transformations)
leaving the gauge sector
diagonal~\cite{Branco:1999nb,Branco:2007nn,EmmanuelCosta:2009bx}.

The Nearest-Neighbour-Interaction (NNI) is an example of a weak basis
in which the up- and down-quark mass matrices, $M_u$ and $M_d$, share
the same texture-zero form:
\begin{equation*}
	M_{u,d}=\begin{pmatrix}
	\;0\; & \;A_{u,d}\; & \;0\; \\
	A^{\prime}_{u,d} & 0 & B_{u,d} \\
	0 & B^{\prime}_{u,d} & C_{u,d}
	\end{pmatrix}\,,
\end{equation*} 
where the constants $A_{u,d}\,$, $A^{\prime}_{u,d}\,$, $B_{u,d}\,$,
$B^{\prime}_{u,d}\,$ and $C_{u,d}$ are independent and complex. Since
this parallel structure is a weak basis, no physical predictions can
be made unless further assumptions are considered. This is the case of
the Fritzsch Ansatz~\cite{Fritzsch:1977vd,Fritzsch:1979zq} where, in
addition of the NNI structure, one requires Hermiticity for both $M_u$
and $M_d$, which cannot be obtained through a weak basis
transformation. It is well known that the Fritzsch Ansatz can no
longer accommodate the current experimental quark mixings. 
However, it was shown in ref.~\cite{Branco:2010tx} that 
deviations from the Hermiticity around 20\% were compatible with the 
experimental data.

It was shown in ref.~\cite{Branco:2010tx} that it is possible to
obtain up- and down-quark mass matrices $M_u$ and $M_d$ with the NNI
structure through the implementation of an Abelian discrete flavour
symmetry in the context of the two Higgs doublet model~(2HDM). In that
context, the minimal realisation is the group $\mathsf{Z}_4$. In a
general 2HDM, a NNI form for each Yukawa coupling matrices cannot be a
weak basis choice. Indeed, the requirement of the
$\mathsf{Z}_4$ symmetry does imply restrictions on the scalar
couplings to the quarks, although one gets no impact on the quark
masses and the Cabibbo-Kobayashi-Maskawa
matrix~\cite{Cabibbo:1963yz,Kobayashi:1973fv}.

The goal of this article is to study whether it is possible to
implement a $\mathsf{Z}_4$ flavour symmetry, as in
refs.~\cite{Branco:2010tx,EmmanuelCosta:2011jq}, that leads to quark
mass matrices $M_u$ and $M_d$ with the NNI form in the context of the
adjoint-$\mathsf{SU(5)}$ Grand Unification~\cite{Perez:2007rm}. The
requirement of a flavour symmetry that enforces a particular pattern
in both up- and down-quark Yukawa couplings has phenomenological
implications for the leptonic sector.  The adjoint-$\mathsf{SU}(5)$
model strengths the weak points left in
ref.~\cite{EmmanuelCosta:2011jq}, namely it improves the
unification of the gauge couplings, solves the mass mismatch between
the charged leptons and down-type quarks at renormalisable level,
alleviates the constraint imposed by the proton lifetime and introduces
a richer mechanism to generate light neutrino masses.

The minimal version of the adjoint-$\mathsf{SU}(5)$
model~\cite{Perez:2007rm} consists in adding to the minimal
$\mathsf{SU}(5)$ version a $\mathsf{45}$ Higgs representation,
$\mathsf{45}_H$, together with an adjoint fermionic field,
$\rho(\mathsf{24})\,$ to generate light neutrino masses. In this
minimal setup one $\rho(\mathsf{24})$ is enough to account for the
observed low-energy neutrino data compatible with a non-zero lightest
neutrino mass~\cite{Dorsner:2006fx, Perez:2007iw, Blanchet:2008cj,
  Kannike:2011fx}. Nevertheless, the nature of the $\mathsf{Z}_4$
symmetry requires an extension of the number of $\rho(\mathsf{24})$
fields, $n_{\mathsf{24}}\,$. The fermionic $\rho(\mathsf{24})$ fields
make possible the generation of neutrino masses through the
interaction with the $\mathsf{45}_H$ field, since gauge interactions
forbid a singlet right-handed neutrino to couple to
$\mathsf{45}_H$. Neutrino masses arise from three different types of
seesaw mechanisms: type-I~\cite{Minkowski:1977sc, Yanagida:1979as,
  GellMann:1980vs, Mohapatra:1979ia}, type-III~\cite{Foot:1988aq,
  Ma:1998dn} and radiative seesaw~\cite{Zee:1980ai,
  Wolfenstein:1980sy, Ma:2006km}. The radiative seesaw is realizable
through the octet-doublet $S_{(8,2)}$ of the $\mathsf{45}_H$ multiplet
at the one-loop level.

This paper is organised as follows: in section~\ref{sec:model} the
$\mathsf{SU}(5) \times \mathsf{Z}_4$ model is described in
detail. Next, in section~\ref{sec:unif} we address the issues of the
unification of the gauge couplings as well as the phenomenology of
proton decay in the model. The successful textures for the effective
neutrino mass matrix together with some comments on the generation of
the baryon asymmetry of the universe through leptogenesis are discussed
in section~\ref{sec:textures}. Then, our numerical results showing the
viability of the leptonic textures considered are sketched in
section~\ref{sec:numerics}. Finally, the conclusions are drawn in
section~\ref{sec:conclusions}.


\section{The model}
\label{sec:model}

The adjoint-$\mathsf{SU}(5)$ model~\cite{Perez:2007rm} contains three
generations of $\mathsf{5^{\ast}}$ and $\mathsf{10}$ fermionic
multiplets which accommodate the SM fermion content. In addition one
adjoint fermionic multiplet $\rho(\mathsf{24})$ is introduced for the
purpose of generating the light neutrino masses and mixings. In this
minimal version, three non-vanishing light neutrino masses arise from
three different seesaw mechanisms, as it will become clear
later. Since our aim is to enlarge the symmetry of the Lagrangian with
an extra $Z_4$ symmetry it forces us to consider $n_{24}$ copies of
$\mathsf{24}$ fermionic field. 

The Higgs sector is composed by an adjoint multiplet,
$\Sigma(\mathsf{24})$, a quintet $\mathsf{5}_H$ and one
$45$-dimensional representation $\mathsf{45}_H$.  Details of the SM
fields contained in the GUT representations are given in
appendix~\ref{reps}. 

The adjoint field $\Sigma$ has the usual role
to break spontaneously the GUT gauge group down to the SM group,
i.e. $\mathsf{SU(3)_{\text{\sc c}} \times SU(2)_{\text{\sc l}} \times
  U(1)_{\text{\sc y}}}$, through its vacuum expectation value (VEV),
\begin{equation}
  \label{eq:vevSM}
  \langle\Sigma\rangle=\frac{\sigma}{\sqrt{60}}\diag(2,2,2,-3,-3)\,.
\end{equation}
The Higgs quintet and the Higgs $\mathsf{45}$-plet give rise to two
doublets, $H_1\in\mathsf{5}_H$ and $H_2\in\mathsf{45}_H$, at
low-energies and two massive $\mathsf{SU}(3)$ colour triplets,
$T_1\in\mathsf{5}_H$ and $T_2\in\mathsf{45}_H$. It is essential that
the triplets $T_{1,2}$ have masses around the unification scale while
the doublets $H_{1,2}$ should remain at the electroweak scale in order
to prevent rapid proton decay - the so called doublet-triplet
splitting problem. The representations $(\bar{3},1,4/3)$ and
$(3,3,-1/3)$ in $\mathsf{45}_H$ could also induce proton
decay. However, it has been shown that the former representation does
not contribute at tree level to proton decay, while some states of the
latter representation contribute with a constraint milder than the one
given by the triplet $T_2\,$~\cite{Dorsner:2012nq}.

Many mechanisms were proposed in order to avoid the
doublet-triplet-splitting problem. One possibility that can be easily
invoked within this framework is the missing partner
mechanism~\cite{Beringer:1900zz,Masiero:1982fe}, which consists in
having the bosonic representations $\mathsf{50}$, $\mathsf{50}^{\ast}$ and
$\mathsf{75}$ instead of the adjoint $\Sigma$ to break the GUT
group. In the missing partner mechanism the scalar doublets are
naturally massless. 

The role of the scalar fields $\mathsf{5}_H$ and $\mathsf{45}_H$ is
then to break the SM group to $SU(3)_{\text{\sc c}}\times
U(1)_{\rm{e.m.}}$ through their VEVs
\begin{equation}
 \label{eq:vev5-45}
  \left\langle\mathsf{5}_H\right\rangle^{\mathsf{T}}\,=\,\left(0,0,0,0,v_{\mathsf{5}}\right)\,,
\end{equation}
and
\begin{equation}
  \left\langle{\mathsf{45}_H}^{\alpha5}_{\beta}\right\rangle\,=\,v_{\mathsf{45}}\left(\delta_{\alpha}^{\beta}-4\,\delta_{4}^{\alpha}
  \delta_{\beta}^{4}\right)\,,\;\alpha,\beta=1,\dots,4\,,
\end{equation}
that are related as
\begin{equation}
  v^2\,\equiv\, \left|v_{\mathsf{5}}\right|^2 + 24\left|v_{\mathsf{45}}\right|^2 =
  \left(\sqrt{2}\,G_F\right)^{-1}= (246.2\,\text{GeV})^2\,,
\end{equation}
where $G_F$ is the Fermi constant. 

Since all representations are well defined, we can specify the nontrivial
transformations of each bosonic/fermionic field $R$ under the
$\mathsf{Z}_4$ flavour symmetry as:
\begin{equation}
  R\longrightarrow\,R^{\prime}=
  e^{i\,\frac{2\pi}{4}\mathcal{Q}(R)}\,R\,,\quad
  \mathcal{Q}(R)\in\mathsf{Z}_4\,.
\end{equation}
The purpose of the discrete symmetry is to obtain the quark mass
matrices, $M_u$, $M_d$, with the NNI form at low energy scales. In
order to preserve the $\mathsf{Z}_4$ symmetry below the unification
scale it is required that $\mathcal{Q}(\Sigma)=0$. In order to
implement the missing partner mechanism, the extra bosonic fields,
$\mathsf{50}$, $\mathsf{50}^{\ast}$ and $\mathsf{75}\,$, must also be
trivial under $\mathsf{Z}_4\,$. Thus, below the unification scale
$\Lambda$ the $\mathsf{Z}_4$ group is preserved in higher orders of
perturbation theory, provided that no Nambu-Goldstone boson appears at
tree-level due to an accidental global
symmetry~\cite{Georgi:1974au}. At low energies one obtains a two Higgs
doublet model with extra fermions invariant under $\mathsf{Z}_4$,
which gets broken once the doublets acquire VEVs.

The $\mathsf{Z}_4$-charges are assigned as follows. First we make
the choice that the $\mathsf{45}_H$ couples to the bilinear
$10_3\,10_3\,$, which implies that
\begin{equation}
\label{eq:45H}
\mathcal{Q}(\mathsf{45}_H)\,=\,-2\,\mathcal{Q}(10_3)\,.
\end{equation}
This particular choice does not eliminate any texture on the leptonic
sector obtained when varying the fermionic
$\mathsf{Z}_4$-charges. Thus, the most general fermionic
$\mathsf{Z}_4$-charges that lead to NNI for the quark mass matrices
$M_{u,d}$ are
\begin{equation}
  \begin{aligned}
    \label{eq:su5q}
    \mathcal{Q}(\mathsf{10}_i)&=(3q_3+\phi,\,-q_3-\phi,\,q_3)\,,\\
    \mathcal{Q}(\mathsf{5^{\ast}}_i)&=(q_3+2\phi,\,-3q_3,\,-q_3+\phi)\,,
  \end{aligned}
\end{equation}
where $\phi\equiv\mathcal{Q}(\mathsf{5}_H)\,$ and
$q_3\equiv\mathcal{Q}(\mathsf{10}_3)$. The charges for the $n_{24}$
adjoint fermions are left free and only some combination of them will
lead to realistic effective neutrino mass matrices, as we will see.

In this model, the most general Yukawa interactions are given by the
following terms:
\begin{equation}
  \label{eq:yukawa}
  \begin{split}
    -& \mathcal{L}_\text{Y}
    =\epsilon_{\alpha\beta\gamma\delta\xi}\left[\left(\Gamma^1_u\right)_{ij}\mathsf{10}^{\alpha\beta}_i\mathsf{10}^{
        \gamma\delta}_j\left(\mathsf{5}_H\right)^{\xi}
      +\left(\Gamma^2_u\right)_{ij}\mathsf{10}^{\alpha\beta}_i\, \mathsf{10}^{\kappa\gamma}_j\left(\mathsf{45}_H\right)^{\delta\xi}_{
        \kappa}\right] \\
    & +
    \left(\Gamma^1_d\right)_{ij}\mathsf{10}^{\alpha\beta}_i\,\mathsf{5}^{\ast}_{j\,\alpha}\left(\mathsf{5}_H^{\ast}\right)_{\beta}
    +
    \left(\Gamma^2_d\right)_{ij}\mathsf{10}^{\alpha\beta}_i\,\mathsf{5}^{\ast}_{j\,\gamma}\left(\mathsf{45}_H^{\ast}\right)^{\gamma
    }_{\alpha\beta}
    +\mathbf{M}^{}_{kl}\tr\left(\rho^{}_k\,\rho^{}_l\right)\\
    &+\boldsymbol\lambda^{}_{kl}\tr\left(\rho^{}_k\,\rho^{}_l\,\Sigma\right)
    +\left(\Gamma^1_{\nu}\right)_{ik}\mathsf{5}^{\ast}_{i\,\alpha}\left(\rho^{}_k\right)^{\alpha}_{\beta}
    \left(\mathsf{5}_H\right)^{\beta}+
    \left(\Gamma^2_{\nu}\right)_{ik}\mathsf{5}^{\ast}_{i\,\alpha}\left(\rho^{}_k\right)^{\gamma}_{\beta}
    \left(\mathsf{45}_H\right)^{\alpha\beta}_{\gamma}+ \text{H.c.}\,,
  \end{split}
\end{equation}
where $\alpha,\beta,\dots=1,\dots,5$ are $\mathsf{SU(5)}$ indices,
$i, j$ are generation indices and $k,l=1,\dots,n_{24}\,$. Notice that
the Yukawa matrix $\Gamma^1_u$ and $\boldsymbol\lambda$ as well as the
mass matrix $\mathbf{M}$ are symmetric while $\Gamma^2_u$ is
antisymmetric. Taking into account the charges given in
eq.~\eqref{eq:su5q}, the Yukawa coupling matrices $\Gamma^{1,2}_{u,d}$
are given by,
\begin{subequations}
  \label{eq:Y}
  \begin{align}
\quad
    \Gamma^1_{u}=
    \begin{pmatrix}
      \;0\; & \;0\; & \;0\;\\
      0 & 0 & b_{u}\\
      0 & {b}_{u}& 0
    \end{pmatrix}
    \,,\quad \Gamma^2_{u}&=
    \begin{pmatrix}
     \;0\; & \;a_{u}\; & \;0\;\\
      a^{\prime}_{u} & 0 &0\\
      0 & 0 & c_{u}
    \end{pmatrix}\,,\\[2mm]
    \Gamma^1_{d}=
    \begin{pmatrix}
      \;0\; & \;a_{d}\; & \;0\;\\
      a^{\prime}_{d} & 0 &0\\
      0 & 0 & c_{d}
    \end{pmatrix}
    \,,\quad \Gamma^2_{d}&=
\begin{pmatrix}
      \;0\; & \;0\; & \;0\;\\
      0 & 0 & b_{d}\\
      0 & b^{\prime}_{d}& 0
    \end{pmatrix}\,.
  \end{align}
\end{subequations}
The up- and down-quark masses as well as the
charged lepton masses are given by 
\begin{equation}
  \label{eq:masses}
  \begin{aligned}
    M_u&=4\,v_{\mathsf{5}}\,\Gamma^1_u\,+8\,v_{\mathsf{45}}\,\Gamma^2_u\,,\\
    M_d&=\,v^{\ast}_{\mathsf{5}}\,\Gamma^1_d+2\,\,v^{\ast}_{\mathsf{45}}\,\Gamma^2_d\,,\\
    M_e&=v^{\ast}_{\mathsf{5}}\,{\Gamma^1_d}^{\mathsf{T}}-6\,v^{\ast}_{\mathsf{45}}\,{\Gamma^2_d}^{\mathsf{T}}\,.
  \end{aligned}
\end{equation}
Substituting the eqs.~\eqref{eq:Y} in the mass matrices given in
eqs.~\eqref{eq:masses} one concludes that all the matrices $M_{u,d,e}$
share the NNI structure. The up-quark mass matrix $M_u$ is no
longer symmetric and the mismatch between the down-type and charged
lepton matrices is now explained as:
\begin{equation}
  M_d-M_e^{T}=8\,v^{\ast}_{\mathsf{45}}\,{\Gamma^2_d}\,.
\end{equation}

The mass matrices of the fermions $\rho_0$, $\rho_3$ and $\rho_8$ arising
from the fermionic-$\mathsf{24}$ fields are given by:
\begin{equation}
  \label{eq:romasses}
  \begin{aligned}
    \mathbf{M}_{0} & = \frac{1}{4}\left( \mathbf{M}\,-\,\frac{\sigma}{\sqrt{30}} \,\boldsymbol\lambda\right)\,,\\
    \mathbf{M}_{3} & = \frac{1}{4}\left( \mathbf{M}\,-\,\frac{3\,\sigma}{\sqrt{30}} \,\boldsymbol\lambda\right)\,,\\
    \mathbf{M}_{8} & = \frac{1}{4}\left( \mathbf{M}\,+\,\frac{2\,\sigma}{\sqrt{30}} \,\boldsymbol\lambda\right)\,.
  \end{aligned}
\end{equation}
Due to the fact that the Higgs field $\Sigma$ is trivial under
$\mathsf{Z}_4$, the matrices $\mathbf{M}$ and $\boldsymbol\lambda$
share the same form and this is also valid for the Majorana matrices
$\mathbf{M}_{0,3,8}$. From the Yukawa interactions written in
eq.~\eqref{eq:yukawa}, one can infer the Yukawa couplings for the
$\rho_0\,$, $\rho_3$ and $\rho_8$ fermion fields, which are then given
by
\begin{equation}
  \label{eq:yukneutrinosro}
  \begin{split}
    -\mathcal{L}_\text{Y} =&
    \frac{\sqrt{15}}{2\sqrt{2}}\left[
      \frac{\cos\alpha}{5}\left(\Gamma^1_{\nu}\right)_{kl}
      \,+\, \sin\alpha\left(\Gamma^2_{\nu}\right)_{kl}
      \right]\,l^{\mathsf{T}}_k i\sigma_2\,{\rho_0}_l\,H\\[2mm]    
    +& \frac{\sqrt{15}}{2\sqrt{2}}\left[
      -\frac{\sin\alpha}{5}\left(\Gamma^1_{\nu}\right)_{kl}
    \,+\cos\alpha\left(\Gamma^2_{\nu}\right)_{kl}
    \right]\,l^{\mathsf{T}}_k i\sigma_2\,{\rho_0}_l\,H^\prime\\[2mm]  
    +&\frac{1}{\sqrt{2}}\left[\cos\alpha\left(\Gamma^1_{\nu}\right)_{kl}\,-3\sin\alpha\left(\Gamma^2_{\nu}\right)_{kl}
      \right]\,l^{\mathsf{T}}_ki\sigma_2\,{\rho_3}_l\,H\\[2mm]    
    -&\frac{1}{\sqrt{2}}\left[\sin\alpha\left(\Gamma^1_{\nu}\right)_{kl}\,+\,3\cos\alpha\left(\Gamma^2_{\nu}\right)_{kl}
      \right]\,l^{\mathsf{T}}_ki\sigma_2\,{\rho_3}_l\,H^\prime\\[2mm] 
    -&\frac{\left(\Gamma^2_{\nu}\right)_{kl}}{\sqrt{2}}
    \,l^{\mathsf{T}}_k i\sigma_2\tr\left(S_{(8,2)}{\rho_8}_l\right)\,,
  \end{split}
\end{equation}
where $S_{(8,2)}$ is the scalar octet-doublet belonging to the
$\mathsf{45}_H$ representation and the doublet space
$\left(H_1,H_2\right)^{\mathsf{T}}$ has also been rotated in terms of new
doublets $\left(H,H^{\prime}\right)^{\mathsf{T}}$ such that $\langle
H\rangle\,=\,v$ and $\langle H^{\prime}\rangle\,=\,0\,$ by the
appropriate transformation:
\begin{equation}
  \begin{pmatrix}
    H \\  H^{\prime}
  \end{pmatrix}=
  \begin{pmatrix}
    \cos \alpha & \sin \alpha \\
    -\sin \alpha & \cos \alpha
  \end{pmatrix}\begin{pmatrix}
    H_1  \\ H_2
  \end{pmatrix}\,,
\end{equation}
with $\tan\alpha\,\equiv\,v_{\mathsf{45}}/v_{\mathsf{5}}\,$.

\begin{figure}[t]
  \begin{center}
    \subfigure[fnm1][Type-I seesaw]{  
      \begin{tikzpicture}[scale=1]
        \draw [fermion] (0,0)--(1,0);
        \draw [fermion] (4,0) -- (3,0);
        \draw [fermion] (1,0) -- (2,0);
        \draw [fermion] (3,0) -- (2,0);
         \draw [scalar] (1,0) --(1,2);
          \draw [scalar] (3,0) --(3,2);
        \node [cross out] at (2,0) {$\boldsymbol\times$};
         \node  at (0.3,0.2) {\scalebox{0.93}{$\nu_i$}};
         \node  at (3.7,0.2) {\scalebox{0.93}{$\nu_j$}};
        \node  at (0.5,1.6) {\scalebox{0.8}{$\langle H\rangle$}};
        \node  at (0.5,1.6) {\scalebox{0.8}{$\langle H\rangle$}};
        \node  at (3.5,1.6) {\scalebox{0.8}{$\langle H\rangle$}};
        \node  at (2.0,-0.4) {\scalebox{0.8}{${\rho_0}_k$}};
      \end{tikzpicture}
    }
    \qquad
    \subfigure[fnm2][Type-III seesaw]{  \begin{tikzpicture}[scale=1]
      \draw [fermion] (0,0)--(1,0);
      \draw [fermion] (4,0) -- (3,0);
      \draw [fermion] (1,0) -- (2,0);
      \draw [fermion] (3,0) -- (2,0);
       \draw [scalar] (1,0) --(1,2);
       \draw [scalar] (3,0) --(3,2);
       \node [cross out] at (2,0) {$\boldsymbol\times$};
       \node  at (0.3,0.2) {\scalebox{0.93}{$\nu_i$}};
         \node  at (3.7,0.2) {\scalebox{0.93}{$\nu_j$}};
       \node  at (0.5,1.6) {\scalebox{0.8}{$\langle H\rangle$}};
        \node  at (3.5,1.6) {\scalebox{0.8}{$\langle H\rangle$}};
       \node  at (2.0,-0.4) {\scalebox{0.8}{${\rho_3}_k$}};
      \end{tikzpicture}  
    }
    \qquad
    \subfigure[fnm3][Radiative seesaw at one-loop level]{  \begin{tikzpicture}[scale=1]
        \draw [fermion] (0,0)--(1,0);
        \draw [scalar] (1,0) arc (180:0:1);
        \draw [fermion] (4,0) -- (3,0);
        \draw [fermion] (1,0) -- (2,0);
        \draw [fermion] (3,0) -- (2,0);
        \draw [scalar] (2,1) --(1.3,2);
        \draw [scalar] (2,1) --(2.7,2);
        \node [cross out] at (2,0) {$\boldsymbol\times$};
        \node  at (0.3,0.2) {\scalebox{0.93}{$\nu_i$}};
        \node  at (3.7,0.2) {\scalebox{0.93}{$\nu_j$}};
        \node  at (1.0,1.6) {\scalebox{0.8}{$\langle H\rangle$}};
        \node  at (3,1.6) {\scalebox{0.8}{$\langle H\rangle$}};
        \node  at (2.0,-0.4) {\scalebox{0.8}{${\rho_8}_k$}};     
        \node  at (2.0,0.6) {\scalebox{0.8}{$S_{(8,2)}$}};    
  \end{tikzpicture}}
  \end{center}
  \caption{\label{fig1} Different seesaw mechanisms present in the model.}
\end{figure}
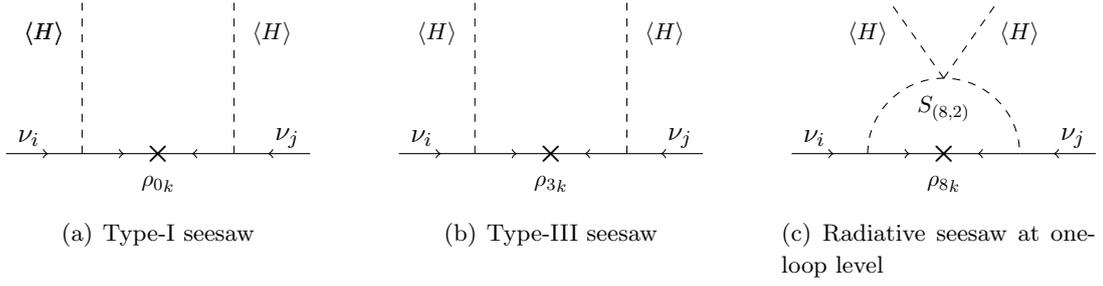

Taking into account the Majorana mass matrices $\mathbf{M}_{0,3,8}$
and the Yukawa interactions given by eq.~\eqref{eq:yukneutrinosro} one
can derive, through the seesaw mechanism, the effective neutrino mass
matrix $m_{\nu}$ which receives three different contributions, as 
drawn in figure~\ref{fig1}.  One has type-I
seesaw~\cite{Minkowski:1977sc, Yanagida:1979as, GellMann:1980vs,
  Mohapatra:1979ia} mediated by the fermionic singlets ${\rho_0}_k$ and
type-III seesaw~\cite{Foot:1988aq,Ma:1998dn} via the exchange of the
$\mathsf{SU(2)}$-triplets ${\rho_3}_k$. There is still the possibility of
generating neutrino masses through the radiative
seesaw~\cite{Zee:1980ai,Wolfenstein:1980sy}, that involves at 1-loop
the fermionic ${\rho_8}_k$ and the scalar doublet-octet $S_{(8,2)}$ present in the
$\mathsf{45}_H$. The neutrino mass matrix obtained after integrating
out the fields responsible for the seesaw mechanism reads as
\begin{equation}
  \label{eq:seesaw}
  \begin{split}
  \left(m_{\nu}\right)_{ij} &= 
  -\,\left(m^D_0\mathbf{M}_0^{\text{-}1}\,{m^D_0}^{\mathsf{T}}\right)_{ij}
  -\,\left(m^D_3\mathbf{M}_3^{-1}\,{m^D_3}^{\mathsf{T}}\right)_{ij}\\[2mm]
  &-\,\frac{v^2\,\zeta}{8\pi^2} \sum_{k=1}^{n_{24}}
  \frac{\left(U_8\,\Gamma^2_{\nu}\right)_{ik}\,\left(U_8\,\Gamma^2_{\nu}\right)_{jk}}{\widetilde{M}_{8\;k}}
  \,\,F\left[\frac{M_{S_{(8,2)}}}{\widetilde{M}_{8\;k}}\right]\,,
  \end{split}
\end{equation}
where $m^D_0\,$, $m^D_3$ are given by
\begin{equation}
  \label{eq:h}
  \begin{aligned}
 m^D_0&=\frac{\sqrt{15}v}{2\sqrt{2}}\left(\frac{\cos\alpha}{5}\,\Gamma^1_{\nu}+\sin\alpha\,\Gamma^2_{\nu}\right)\,,\\
 m^D_3&=\frac{v}{\sqrt{2}}\left(\cos\alpha\,\Gamma^1_{\nu}-3\sin\alpha\,\Gamma^2_{\nu}\right)\,,
  \end{aligned}
\end{equation}
and $\widetilde{M}_{8\;k}$ are simply the mass-eigenvalues of the
Majorana matrix $\mathbf{M}_8\,$. The unitary matrix $U_8$ does the
rotation of the yukawa matrix $\Gamma^2_{\nu}$ to the basis where the
matrix $\mathbf{M}_8$ is diagonal. The coefficient $\zeta$ is a linear
combination of the coefficients in the Higgs potential terms given in
eqs.~\eqref{eq:pot45} and~\eqref{eq:pot45e5} in the
appendix~\eqref{sec:pot}. The loop function $F(x)$ is given by
\begin{equation}
F(x)\,\equiv\,\frac{x^2-1-\log x}{(1-x^2)^2}\,.
\end{equation}
If one assumes
$\widetilde{M}_{8\;k}\gg\widetilde{M}_{0\;k}>\widetilde{M}_{3\;k}$
then it suppresses the 1-loop radiative seesaw contribution.

Before closing this section, it is important to comment about the
Higgs potential. The most general Higgs potential is given explicitly
in the appendix~\eqref{sec:pot}.  Notice that, terms involving
simultaneously the fields $\mathsf{5}_H$, $\mathsf{24}_H$ and
$\mathsf{45}_H$ are forbidden by the $\mathsf{Z}_4\,$ symmetry.  This
gives rise to an accidental global continuous symmetry which, upon
spontaneous electroweak symmetry breaking, would lead to a massless
Nambu-Goldstone boson at tree level~\cite{Weinberg:1972fn}.  A simple
way to cure this problem is by adding a complex $\mathsf{SU}(5)$ Higgs
singlet $S$ non-trivially charged under $\mathsf{Z}_4\,$,
i.e. $\mathcal{Q}(S)=-2\,q_3-\phi\,$, where its potential is given by
\begin{equation}
\label{eq:potS}
  V_{S}\,=\,\left( \lambda_{\text{\sc sb}}\, \mathsf{5}^{\ast}_{\alpha}\,
  24^{\gamma}_{\beta}\, 45^{\alpha \beta}_{\gamma}\,S
  \,+\,\text{H.c.}\right)\,+\,\frac{1}{2}\,\mu_S^2\,|S|^2
  \,+\,\lambda_S\,|S|^4\,+\,\lambda^{\prime}_S\,(S^4
  \,+\,\text{H.c.})\,,
\end{equation} 
and leads at low-energy to an effective interaction, once the scalar
$S$ acquires vacuum expectation value,
\begin{equation}
\label{eq:mu12}
\lambda_{\text{\sc sb}}\,\sigma\,\langle S\rangle\,H^{\dagger}_1\,H_2\,+\,\text{H.c.}\,,
\end{equation}
which softly breaks the symmetry $\mathsf{Z}_4\,$.

\section{Unification and Proton Stability}
\label{sec:unif}

According to the previous section, between the unification scale and
$M_Z=91.1876\pm0.0021\,\text{GeV}$ scale~\cite{Beringer:1900zz} one
has a 2HDM with extra fermions, namely ${\rho_0}_k\,$, ${\rho_3}_k\,$,
and ${\rho_8}_k\,$, and the two scalars $\Sigma_3$ and $\Sigma_8$ that
can have lower masses. The coulored triplets, $T_1$ and $T_2\,$, and
the other scalars contained in the $\mathsf{45}_H$ are set their
masses arround the GUT scale. We also assume $M_{\Sigma_3} \simeq
M_{\Sigma_8}$. The running of the three gauge coupling constants
$\alpha_i \, (i=1,2,3)$ in the 2HDM with extra particle content can be
obtained easily at the one-loop level as
\begin{subequations}
\begin{align}
\label{eq:RGEsol1}
\alpha^{-1}_1(\mu)&=\alpha^{-1}_1(M_Z)-\frac{b_1}{2\pi}\log\left(\frac{\mu}{M_Z}\right)
-\sum_I\frac{b^I_1}{2\pi}\log\left(\frac{\mu}{M_I}\right)
\,,\\
\label{eq:RGEsol2}
\alpha^{-1}_2(\mu)&=\alpha^{-1}_2(M_Z)-\frac{b_2}{2\pi}\log\left(\frac{\mu}{M_Z}\right)
-\sum_I\frac{b^I_2}{2\pi}\log\left(\frac{\mu}{M_I}\right)
\,,\\
\label{eq:RGEsol3}
\alpha^{-1}_3(\mu)&=\alpha^{-1}_3(M_Z)-\frac{b_3}{2\pi}\log\left(\frac{\mu}{M_Z}\right)
-\sum_I\frac{b^I_3}{2\pi}\log\left(\frac{\mu}{M_I}\right)
\,,
\end{align}
\end{subequations}
where $\alpha_1 = 5/3\,\alpha_y,\, \alpha_2 = \alpha_w$ and $\alpha_3
= \alpha_s$; the $b_i$ constants are the usual one-loop beta
coefficients corresponding to the 2HDM, listed in
section~\ref{reps}. $M_I$ denotes an intermediate energy scale for extra particle $I$ between
the electroweak scale $M_Z$ and the GUT scale $\Lambda$, and the
coefficients $b_{i}^I$ account for the new contribution to the
one-loop beta functions $b_{i}$ above the threshold $M_I$. At the
unification scale $\Lambda$, the gauge couplings $\alpha_i$ obey to the
relation
\begin{equation}\alpha^{}_U \,\equiv\, \alpha_1 (\Lambda)= \alpha_2(\Lambda)= \alpha_3(\Lambda)\,.
\end{equation}

To get some insight into the unification in the one-loop approximation, let us
define the effective beta coefficients $B_i$~\cite{Giveon:1991zm},
\begin{equation}
B_i\equiv b_i+\sum_I b_i^I\,r_I,
\end{equation}
where the ratios $0\leq r_I\leq1$ that takes into account the intermediate scales
are given by
\begin{equation}
r_I = \frac{\ln\left(\Lambda/M_I\right)}{\ln\left(\Lambda/M_Z\right)}\,.
\end{equation}
It is also convenient to introduce the
differences $B_{ij}\equiv B_i-B_j$, define as
\begin{equation}
B_{ij}= B^{\text{2HDM}}_{ij}+\sum_I\Delta^I_{ij}r_I\,,
\end{equation}
where $B^{\text{2HDM}}_{ij}$ corresponds to the 2HDM particle contribution and
\begin{equation}
\Delta^I_{ij}\equiv b^I_i-b^I_j\,.
\end{equation}
The following $B$-test is then obtained,

\begin{equation}  \label{eq:Btest}
B\equiv\frac{B_{23}}{B_{12}}=\frac{\sin^2\theta_W-\dfrac{\alpha}
{\alpha_s}}
{\dfrac{3}{5}-\dfrac{8}{5}\sin^2\theta_W}\,,
\end{equation}
together with the GUT scale relation
\begin{equation} \label{eq:Ltest}
B_{12}\, \ln \left(\frac{\Lambda}{M_Z}\right)= \frac{2\pi}{5\alpha}\left(3-8\sin^2\theta_W\right).
\end{equation}

Notice that the right-hand sides of eqs.~\eqref{eq:Btest} and~\eqref{eq:Ltest}
depend only on low-energy electroweak data. using
the following experimental values at $M_Z$~\cite{Beringer:1900zz}
\begin{align}
\alpha^{-1}&=127.916\pm0.015\,, \\
\sin^2\theta_W&=0.23116\pm0.00012\,, \\
\alpha_s&=0.1184\pm0.0007\,,
\end{align}
the above relations read as
\begin{align}\label{eq:Btestexp}
\begin{split}
B&=0.718\pm0.003\,, \\
B_{12}\,\ln\left(\frac{\Lambda}{M_Z}\right)&=185.0\pm0.2\,.
\end{split}
\end{align}

The coefficients $B_{ij}$ that appear in the left-hand sides of
eqs.~\eqref{eq:Btest} and~\eqref{eq:Ltest} strongly depend on the particle
content of the theory. For instance, considering the SM particles with $n_H$
light Higgs doublets, one has $b_1=20/5+n_H/10$, $b_2=-10/3+n_H/6$ and $b_3=-7$,
so that these coefficients are given by
\begin{equation}  \label{eq:beffSM}
B_{12}=36/5\,,\quad B_{23}=4\,.
\end{equation}
where $n_H=2$ is set and $B=5/9$ is then not compatible with the
calculated value in eq.~\eqref{eq:Btestexp} and clearly, the B-test
fails badly in the 2HDM case, so that extra particles are needed.  In
table~\ref{tab1} we present the relevant contributions $\Delta_{ij}$
to the $B_{ij}$ coefficients of our setup which include, besides the
2HDM threshold, the fermions ${\rho_0}_k\,$, ${\rho_3}_k\,$, and
${\rho_8}_k\,$, and the two scalars $\Sigma_3$ and $\Sigma_8$ are
considered. For simplicity, we assume the remaining particles at the
unification scale and therefore they do not contribute to the gauge
coupling running.

\begin{table}[h]
\caption{\label{tab1} The relevant $\Delta_{ij}$ contributions to the $B_{ij}$
coefficients in the $\mathsf{SU}(5)\times\mathsf{Z}_4$ model.}
\centering
\begin{tabular}{cccccc}
\\
\hline\hline
& 2HDM & $\rho_3$ & $\rho_8$ & $\Sigma_3$ & $\Sigma_8$ \\ \hline
$\Delta_{12}$ & 36/5 & -4/3 & 0  & -2/3 & 0  \\
$\Delta_{23}$ &  4 & 4/3 & -2 & 2/3 & -1 \\
\hline\hline
\end{tabular}
\end{table}

\begin{figure}[t]
  \begin{center}
    \includegraphics[width=14cm,clip]{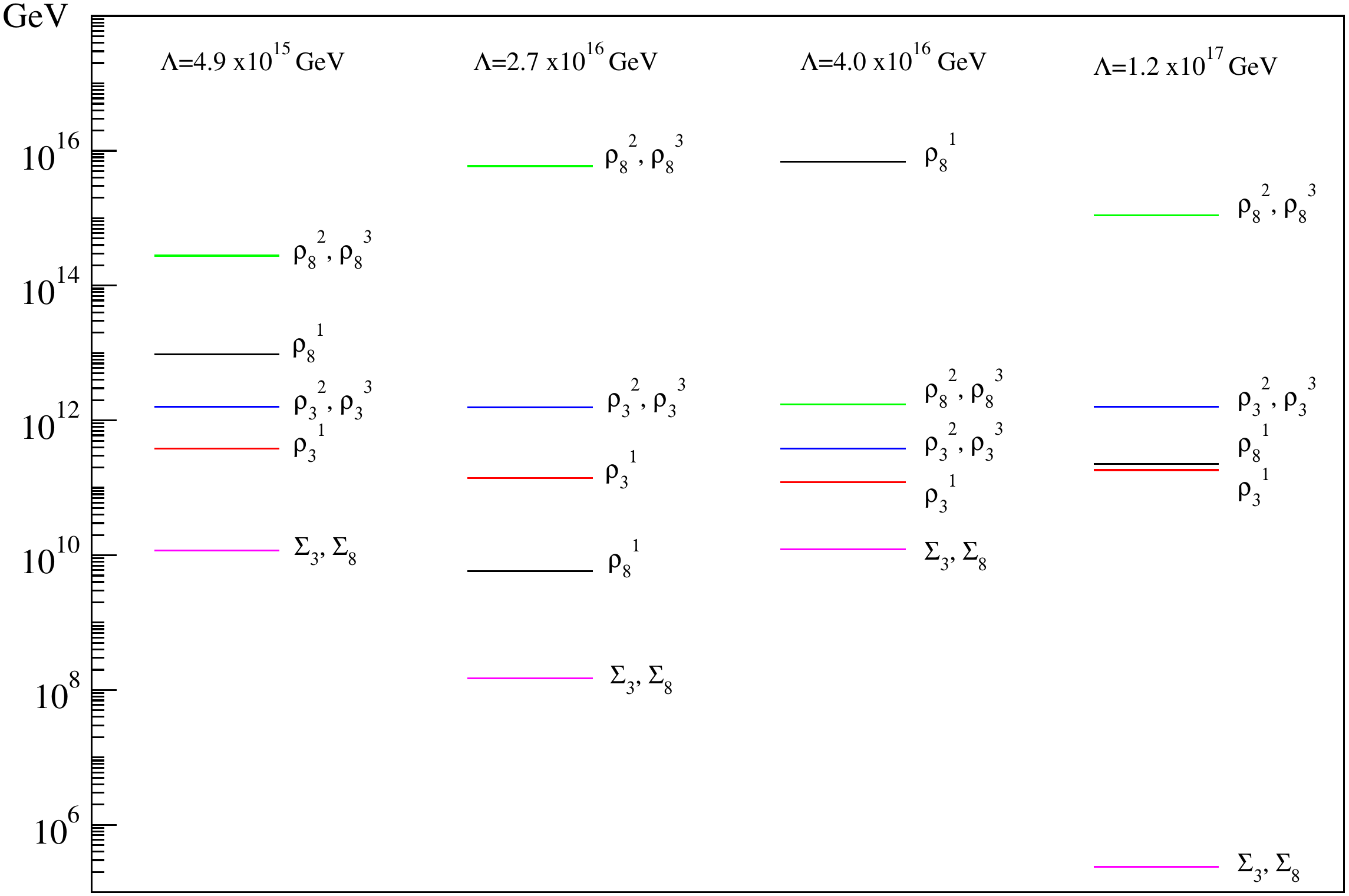}
  \end{center}
  \caption{\label{fig:unif} Four illustrative examples showing the mass
    spectrum of the adjoint fermionic fields, $\Sigma_3$ and
    $\Sigma_8$ for different unification scales $\Lambda$.}
\end{figure}

Notice that eqs.~\eqref{eq:Btestexp} require $\Delta^I_{12}<0$ and
$\Delta^I_{23}>0$, it becomes clear from table~\ref{tab1} all extra particles
considered in the running improve the unification. We have scanned
different ratios $r_I$ and we obtained a large range of solutions that
lead to a perfect unification within the experimental errors. The fact
that more than one adjoint fermionic field $\rho_k$ is present, it
improved the range of intermediate scales $M_I$ consistent with
unification. It is now difficult to find strong correlations among the
intermediate scales $M_I$. In addition, the possible values for the
unification scale $\Lambda$ can vary many orders of magnitude. For
illustration, in figure~\ref{fig:unif} we have drawn the mass spectrum
of the extra particles included in the running. All the solution
obtained are in agreement with a unified gauge coupling
$g_U=\sqrt{4\pi/\alpha_U^{-1}}<1$, where in our numerics we obtained $\alpha_U^{-1}\approx37\,$.

Concerning the proton decay some comments are in order. In this model
there are mainly two different sources for proton decay, namely via
the exchange of the lepto-quark gauge bosons $X,Y$ or via coulored
Higgs triplets. Proton decay in both scenarios are mediated four
fermion interactions (dimension-six operators).

The gauge bosons $X,Y$ become massive through the Higgs mechanism with
a common mass, $M_V$,
\begin{equation}
M_V=\frac{25}{8}g_U^2\sigma^2\,.
\end{equation} To suppress the $X,Y$ boson
proton decay channels, one has necessarily that $M_V\gg m_p$ (the
proton mass), that leads to the  estimation of the proton decay width 
as~\cite{Langacker:1980js}:
\begin{equation}
\Gamma\approx \alpha_U^2 \frac{m_p^5}{M_V^4}\,.
\end{equation}
Making use of the most restrictive constraints on the partial proton
lifetime $\tau(p\rightarrow \pi^0 e^+)>8.2\times10^{33}$
years~\cite{Beringer:1900zz}, one can derive a rough lower bound for
the $X,Y$ mass scale $M_V$,
\begin{equation}
\label{eq:lower}
M_V>4.1\times10^{15}\,\text{GeV}\,,
\end{equation}
which corresponds a $\alpha_U^{-1}\approx 37\,$. Since we assume for the unification scale
$\Lambda\sim M_V$, the constraint given by eq.~\eqref{eq:lower}
determines the scale where the gauge couplings should unify (for a
recent review see~\cite{Nath:2006ut}).

The proton decay through the exchange of Higgs colour triplets
$T_1,T_2$ is very suppressed, since their suppression is proportional
to products of Yukawa couplings, and therefore they are much smaller
than the gauge couplings. Indeed the contribution of these
dimension-six operators vanishes at tree-level when the $\mathsf{Z}_4$
symmetry is exact~\cite{EmmanuelCosta:2011jq}. The dimension-six
operators contributing to the proton decay via the colour triplet
exchange are given at tree-level by:
\begin{equation}
\label{eq:pdecay}
\sum_{n=1,2}
\frac{\left(\Gamma^n_u\right)_{ij}\left(\Gamma^n_d\right)_{kl}}{M^2_{T_n}}
\left[\frac{1}{2} (Q_iQ_j)(Q_kL_l)+(u^c_ie^c_j)(u^c_kd^c_l)\right]\,.
\end{equation}
It is then clear from the pattern of the Yukawa coupling matrices
$\Gamma^{1}_{u}$ and $\Gamma^{2}_{d}$ given in eqs.~\eqref{eq:Y} that
the only possible non-vanishing contribution of the dimension-six
operators given in eq.~\eqref{eq:pdecay} involve necessarily fermions
of the third generation. One concludes that at tree-level the proton
does not decay through the four-fermion interactions described by the
operators given in eq.~\eqref{eq:pdecay}.

\section{Effective Neutrino Textures}
\label{sec:textures}

The flavour symmetry present in our model constrains the charges of
the fermion fields to be of the form in eq.~\eqref{eq:su5q}. Such
charge assignment does not imply any restriction in the neutrino
sector. Hence the charges of the 24 dimensional fermionic
representations responsible for the neutrino masses are free. In
order to analyse the possible patterns for the effective neutrino mass
matrices, $m_\nu$, we have considered all the possible values for the
$\mathsf{Z_4}$ charges of the adjoint fermionic fields.

Searching for the minimal $\mathsf{SU}(5)\times\mathsf{Z}_4$ model,
i.e. the model with the minimal matter content, we have started by the
possibility of having only one extra 24 fermionic representation, as
in the $\mathsf{SU(5)}$ adjoint original
model~\cite{Perez:2007rm}. However, given the particularities of the
$\mathsf{Z_4}$ symmetry, the neutrino mixing pattern that emerges from
this picture is now not consistent with the experimental neutrino
data~\cite{Tortola:2012te}. Adding a second $\mathsf{24}$ fermionic
representation does not solve the problem and again the predicted
neutrino mixing angles are not in agreement with neutrino
oscillations.

The situation changes when we consider three $\mathsf{24}$ fermionic
representations. In this case we obtain different possibilities for
the light neutrino mass matrix $m_\nu$ that coincide with the matrices
found in ref.~\cite{EmmanuelCosta:2011jq}, where a similar
$\mathsf{Z_4}$ flavour symmetry was imposed in the context of three
right-handed neutrinos.  From the various textures for the effective
neutrino mass matrix $m_{\nu}$ found in the scan, only two solutions
can account successfully for the low-energy neutrino data, namely
\begin{equation}
  \label{eq:three}
    m^A_{\nu}\,=\,\begin{pmatrix}
      \;0\;& \;\ast\; & \;0\; \\
       \ast & \ast & \ast\\
        0 & \ast & \ast
      \end{pmatrix}
      \qquad \text{and}\qquad
    m^{A_{\scriptscriptstyle(12)}}_{\nu}\,=\,\begin{pmatrix}
     \;\ast\; & \;\ast\; & \;\ast\; \\
     \ast & 0 & 0\\
     \ast & 0 & \ast
    \end{pmatrix}\,,
\end{equation}
where the index $(12)$ refers to the fact that texture-$A_{\scriptscriptstyle(12)}$ is a
permutated form of the texture-$A$ through the permutation matrix $P_{\scriptscriptstyle(12)}\,$,
\begin{equation}
\label{eq:P12}
P_{\scriptscriptstyle(12)}=\begin{pmatrix}
\;0\;& \;1\;& \;0\;\\
1&0&0\\
0&0&1
\end{pmatrix}\,,
\end{equation}
isomorphic to the symmetric group $S_3$. Thus, the effective mass
matrices $m^A_{\nu}$ and $m^{A_{\scriptscriptstyle(12)}}_{\nu}$ are related by:
\begin{equation}
m^{A_{\scriptscriptstyle(12)}}_{\nu}=P_{\scriptscriptstyle(12)}\,m^A_{\nu}\,P^{\mathsf{T}}_{\scriptscriptstyle(12)}\,.
\end{equation}
The compatibility of the above textures with the experimental neutrino
data has been analyzed in detail in ref.~\cite{EmmanuelCosta:2011jq}
where it was found that just the two-zero textures in $A$ and
$A_{\scriptscriptstyle(12)}$ are phenomenologically viable. An
important result is that texture $A$ is compatible only with normal
hierarchy (NH) while texture $A_{12}$ turns to be compatible only with
inverted hierarchy (IH) in the light neutrino mass spectrum.

At this point, we would like to remark that the textures given in
eq.~\eqref{eq:three} have also been studied in the literature
previuosly. According to the standard
terminology~\cite{Frampton:2002yf}, our allowed matrices $A$ and
$A_{\scriptscriptstyle(12)}$ would correspond to the ones labelled as
A$_2$ and D$_1$, respectively. Texture D$_1$ has been shown to be
either disallowed~\cite{Frampton:2002yf,Desai:2002sz,Dev:2006qe} or
very marginally allowed~\cite{Guo:2002ei}. However, all of these
previous works assume a diagonal charged lepton mass matrix while here
we are considering a charged lepton mass matrix with NNI
form. Therefore the results in the literature do not strictly apply to
our case.

Up to here we have only considered the $\mathsf{Z_4}$ charges for the
$\mathsf{24}$ fermionic fields such as the matrices $\mathbf{M}$ and
$\boldsymbol\lambda$ are non-singular, i.e., $|\mathbf{M}| \neq 0$ and
$|\boldsymbol\lambda| \neq 0\,$ and therefore the Majorana matrices
$\mathbf{M}_x,\,x=0,3,8$ are non-singular as well. This condition is
mandatory in order to derive the effective neutrino mass matrix using
the seesaw formula in eq.~\eqref{eq:seesaw}. However, given the
flavour symmetry present in our model, among all the possible
configurations one can have $|\mathbf{M}_{x}| = 0$.  In these cases
$\mathbf{M}_{x}^{-1}$ is not defined and therefore we can not use the
standard seesaw formula, but instead we should consider the singular
seesaw mechanism. The singular seesaw, first suggested in the context
of a GUT framework~\cite{Johnson:1986gt}, has been considered in
relation to different anomalies in neutrino physics such as the
Simpson neutrino~\cite{Glashow:1990dg,Fukugita:1991ae,Allen:1991zc} or
the LSND signal~\cite{Chun:1998qw, Liu:1998qp,
  Chikira:1998qf,Stephenson:2004wv}. In both cases the modified seesaw
scheme has been used to obtain a neutrino mass spectrum with a
singlet/sterile neutrino in the energy range between light neutrino
masses below the eV and the heavy neutrinos at the seesaw scale.

Here we have analysed the neutrino mass spectrum that emerges from the
singular seesaw mechanism with two and three 24 fermionic
representations. According to our calculations, it is not possible to
obtain an effective neutrino mass matrix compatible with experimental 
data in any of the cases, since we always get too few light neutrino
states.

In summary, our analysis shows that the symmetry requires the presence
of at least three fermionic $\mathsf{24}$ representations and no
singular seesaw, i.e., $|\mathbf{M}_{x}| \neq 0$. Since the parameters
of the last matrix depend on the $\mathsf{Z_4}$ neutrino charges
$\mathcal{Q}(\mathsf{24}_i)$, we can conclude that the neutrino
phenomenology has an impact on the $\mathsf{Z_4}$ neutrino charges.

\begin{table}[h]
\centering
\small
\caption{\label{tab:results} The $\mathsf{Z_4}$ fermionic field
charges for phenomenologically
viable effective neutrino textures.}
\begin{tabular}{cccccccc}
\\ \hline \hline \\[-2mm]
$m_\nu$ & $\mathbf{M}_{0,3,8}$ & $\mathcal{Q}(\mathsf{24}_i)$ & $m^D_{0,3}$ &
$\mathcal{Q}(\mathsf{5}^{\ast})$ & $\mathcal{Q}(\mathsf{10}_i)$ & $\mathcal{Q}(\mathsf{5}_H)$ &
$\mathcal{Q}(\mathsf{45}_H)$ \\[2mm]
\hline \hline \\
\multirow{9}{*}{$\begin{pmatrix}
0 & \ast & 0\\
\ast & \ast & \ast\\
0 & \ast & \ast
\end{pmatrix}$} &    \multirow{2}{*}{$\begin{pmatrix}
0 & 0  & \ast\\
0 & \ast & 0\\
\ast & 0 & 0
\end{pmatrix}$} & \multirow{4}{*}{(1,2,3)} &
$\begin{pmatrix}
0 & 0 & \ast\\
\ast & \ast & 0\\
0 & \ast & \ast
\end{pmatrix}$ & (3,1,0) & (0,2,1) & 1 & 2 \\[6mm]
& & &
$\begin{pmatrix}
\ast & 0 & 0\\
0 & \ast & \ast\\
\ast & \ast & 0
\end{pmatrix}$ & (1,3,0) & (0,2,3) & 3 & 2 \\[6mm]
&  \multirow{2}{*}{$\begin{pmatrix}
\ast & 0 & 0\\
0 & 0 & \ast\\
0 & \ast & 0
\end{pmatrix}$} &   \multirow{4}{*}{(0,1,3)} &
$\begin{pmatrix}
0 & \ast & 0\\
\ast & 0 & \ast\\
\ast & \ast & 0
\end{pmatrix}$ & (1,3,2) & (2,0,3) & 1 & 2 \\[6mm]
& & &
$\begin{pmatrix}
0 & 0 & \ast\\
\ast & \ast & 0\\
\ast & 0 & \ast
\end{pmatrix}$ & (3,1,2) & (2,0,1) & 3 & 2 \\  \\
\hline   \\
\multirow{9}{*}{ $\begin{pmatrix}
\ast & \ast & \ast\\
\ast & 0 & 0\\
\ast & 0 & \ast
\end{pmatrix}$ }&    \multirow{2}{*}{$\begin{pmatrix}
0 & 0 & \ast\\
0 & \ast & 0\\
\ast & 0 & 0
\end{pmatrix}$ }
&    \multirow{4}{*}{(1,2,3)}& $\begin{pmatrix}
\ast & \ast & 0\\
0 & 0 & \ast\\
0 & \ast & \ast
\end{pmatrix}$
& (2,0,1) & (1,3,0) &  1 & 0 \\[6mm]
& & &
$\begin{pmatrix}
0 & \ast & \ast\\
\ast & 0 & 0\\
\ast & \ast & 0
\end{pmatrix}$ & (2,0,3) & (3,1,0) &  3 & 0 \\[6mm]
&    \multirow{2}{*}{$\begin{pmatrix}
\ast & 0 & 0\\
0 & 0 & \ast\\
0 & \ast & 0
\end{pmatrix}$} &  \multirow{4}{*}{(0,1,3)} &
$\begin{pmatrix}
\ast & 0 & \ast\\
0 & \ast & 0\\
\ast & \ast & 0
\end{pmatrix}$ &  (0,2,3) & (3,1,2) & 1 & 0 \\[6mm]
& & &
$\begin{pmatrix}
\ast & \ast & 0\\
0 & 0 & \ast\\
\ast & 0 & \ast
\end{pmatrix}$ & (0,2,1) & (1,3,2) & 3 & 0 \\ \\
\hline
\end{tabular}
\end{table}
  
We present in table ~\ref{tab:results} the $\mathsf{Z_4}$ charge
assignment for the fermionic fields that leads to the successful
textures for the effective neutrino mass matrix $A$ and
$A_{\scriptscriptstyle(12)}$, discussed above. The textures for the
matrices $\mathbf{M_x}$ and $m^D_{0,3}$ are also shown. The charges
for the fermionic fields $\mathsf{10}$ and $\mathsf{5}^{\ast}$ as well
as the Higgs fields $\mathsf{5}_H$ and $45_H$ follow the relations in
eqs.~\eqref{eq:45H} and~\eqref{eq:su5q}. It is worth pointing out that all
Dirac neutrino matrices $m^D_{0,3}$ obtained through the scan have
four texture zeroes.

\subsection*{Leptogenesis}
\label{sec:leptogenesis}

In this section we would like to briefly comment about the possibility of
having leptogenesis in our model. 

As already discussed in refs.~\cite{Blanchet:2008cj,Kannike:2011fx},
the out-of-equilibrium decays of the fermionic fields $\rho_0$ and
$\rho_3$ in the 24 fermionic representation may generate an asymmetry
in the leptonic content of the universe.  In the presence of sphaleron
processes this leptonic asymmetry would be partially converted into a
baryon asymmetry, explaining the observed matter-antimatter asymmetry
of the universe. Depending on the mass hierarchy among $\rho_0$ and
$\rho_3$, the main contribution to the leptonic asymmtery will be
dominated by the decays of one of them.  In principle, our model has
enough freedom to have different mass spectra for the fermionic fields
and therefore, in contrast to the results shown in
refs.~\cite{Blanchet:2008cj,Kannike:2011fx}, in our case the leptonic
asymmetry may be generated by the decay of $\rho_0$ ($\epsilon_0$) or
$\rho_3$ ($\epsilon_3$).  In both cases, the expression for the
generated CP asymmetry would be proportional to:
\begin{subequations}
\begin{align}
\label{eq:eps_rho3}
\epsilon_3& \propto \sum_{j\neq1} \imag\left[
  \left(m_3^{D\dagger} m_3^{D}\right)_{1j}^2\right]\,,\\
\label{eq:eps_rho0}
\epsilon_0& \propto \sum_{j\neq1} \imag\left[
  \left(m_0^{D\dagger} m_0^{D}\right)_{1j}^2\right]\,.
\end{align}
\end{subequations}
Despite the specific flavour structure of the $m_i^D$ matrices,
induced by the $\mathsf{Z_4}$ symmetry (see eq.~\eqref{eq:h}), we have
checked that in principle there are no cancelations in the terms above
and therefore the leptonic asymmetry generated by $\rho_0$ and
$\rho_3$ can be different from zero. It is clear that more accurate
predictions about leptogenesis would require further calculations,
considering the effect of the washout over the initial leptonic
asymmetry as well as the dynamical evolution of the asymmetry with the
solution of Boltzmann equations.  For the moment, however, our goal is
just to show that the model presented here has enough freedom in the
choice of masses and couplings so in principle it is possible to
accommodate the CP asymmetry. Further considerations as, for instance,
the constraints on the model coming from the requirement of a baryon
asymmetry consistent with the observations will be discussed
elsewhere.

\section{Numerical Results}
\label{sec:numerics}

In this section we analyse the phenomenological implications of the
effective neutrino mass matrices $m^{A}_\nu$ and
$m^{A_{\scriptscriptstyle(12)}}_\nu$ in eq.~\eqref{eq:three}. Since
the flavour symmetry is valid under perturbative corrections until the
breaking of the electroweak gauge symmetry, the form of the Yukawa
matrices $\Gamma_d^{1,2},\, \Gamma_{\nu}^{1,2}$ and the Majorana mass
matrices $ \mathbf{M}_{0,3,8}$ remains unchanged. Thus, one can
extract the predictions for $M_{e}$ and $m_\nu$ and confront them with
the observed neutrino data at $M_Z$ energy scale. The effective
neutrino mass matrices obtained
$m^{A/A_{\scriptscriptstyle(12)}}_\nu$, as already mentioned, are the
same as those analysed in ref.~\cite{EmmanuelCosta:2011jq}. However,
the new measurements of the reactor mixing angle
$\theta_{13}$~\cite{An:2012eh} have changed the theoretical picture of
the light neutrino mixings since then.  Therefore, it is worth to
revisit the previous analysis in ref.~\cite{EmmanuelCosta:2011jq} to
take into account these new bounds.

Without loss of generality one can write the charged lepton mass
matrix, $M_{e}$, and the effective neutrino mass matrices,
$m^{(g)}_{\nu}$ as:
\begin{subequations}
\label{eq:lep}
\begin{align}
\label{eq:lepell}
&M_{e}=K_{e}\begin{pmatrix}
    \;0\; & \;A_{e}\; & \;0\;\\
    A^{\prime}_{e} & 0 & B_{e}\\
    0 & B^{\prime}_{e} & C_{e}
   \end{pmatrix}\,,\\[3mm]
\label{eq:lepNu}
&m_{\nu}^{g}=P_g\begin{pmatrix}
\,\, 0 \,\, & \,\, A_{\nu}\,\, & 0\\
A_{\nu} & B_{\nu} & C_{\nu} \\
 0 & C_{\nu} & D_{\nu}\,e^{i\varphi}
 \end{pmatrix}P^{\mathsf{T}}_g\,,
\end{align}
\end{subequations}
where the permutation $g=e$ or $(12)$ according to the
table~\ref{tab:results} and the constants
$A_{e,\nu},\,B_{e,\nu},A^{\prime}_{e},\,B^{\prime}_{e},\,C_{e,
  \nu},\,D_{\nu}$ are taken real and positive. The diagonal phase
matrix $K_{e}$ can be parameterised as
\begin{equation}
K_{e}=\diag(e^{i\kappa_1},e^{i\kappa_2},1)\,,
\end{equation}
and the phase $\varphi$ in eq.~\eqref{eq:lepNu} cannot be removed by
any field redefinition. 

Although the number of the parameters encoded in the pair
$M_e,m_{\nu}$ is 12, as the number of independent physical
parameters experimentally observed at low energy, the zero pattern
exhibited in eqs.~\eqref{eq:lep} does imply new constraints among the
independent physical parameters, as it will be shown. The PMNS matrix $U$ is given by
\begin{equation}
U\,=\,O^{\mathsf{T}}_e \,K^{\dagger}_e\,P_{\scriptscriptstyle g}\, U_{\nu}\,,
\end{equation}
where the orthogonal matrix $O_e$ is the one that diagonalises $M_eM^{\dagger}_e$ as
\begin{equation}
  \left(K_e\,O_e\right)^{\dagger}M_e\,M^{\dagger}_e\,\left(K_e\,O_e\right)\,=\,\diag(m_e\,,
  m_{\mu}\,, m_{\tau})\,,
\end{equation}
while the unitary matrix $U_{\nu}$ diagonalises $m_{\nu}$ as
\begin{equation}
  \left(P_g\,U_{\nu}\right)^{\mathsf{T}}\,m^g_{\nu}\,\left(P_g\,U_{\nu}\right)\,=\,\diag(m_1\,,m_2\,,m_3)\,.
\end{equation}

The knowledge of the low-energy neutrino mixings appears in the
literature in terms of the parameters $\theta_{12}\,, \,\theta_{13}\,,
\,\theta_{23}\,$ and $\delta$ of the Standard Parametrisation
(SP)~\cite{Beringer:1900zz}, defined in terms of PMNS matrix
$U$ invariants as
\begin{equation}
\begin{aligned}
\sin\theta_{12}& \equiv\frac{|U_{e2}|^2}{\sqrt{1-|U_{e3}|^2}}\,,\\
\sin\theta_{13}& \equiv|U_{e3}|\,,\\
\sin\theta_{23}& \equiv\frac{|U_{\mu3}|^2}{\sqrt{1-|U_{e3}|^2}}\,,
\end{aligned}
\end{equation}
and the phase $\delta$ is given by the Dirac-phase invariant, $I$,
\begin{equation}
I \equiv \imag\left(U_{\mu3} U^{\ast}_{e3} U_{e2} U^{\ast}_{\mu2}\right)
=\frac18\cos\theta_{13}\, \sin2\theta_{12}\, \sin2\theta_{23}\,
\sin2\theta_{13}\sin\delta\,.
\end{equation}
Due to the fact that the PMNS matrix is not rephasing invariant on the
right, one defines the Majorana-type phases, $\varphi^{\alpha}_{ij},$  free of any kind of
parametrisation as~\cite{Branco:2008ai}:
\begin{equation}
\label{eq:MajoranaPhases}
\varphi^{\alpha}_{ij} \equiv \arg \left(U_{\alpha i}\,U^{\ast}_{\alpha j} \right)\,.
\end{equation}
It has been shown in ref.~\cite{Branco:2008ai} that the PMNS matrix 
can be fully reconstructed by six independent Majorana-type phases
from eq.~\eqref{eq:MajoranaPhases} taking into account that $U$
is a unitary matrix. The Dirac-type phase $\delta$ can therefore be
expressed as the difference of two Majorana-type phases:
\begin{equation}
I=\left|U_{\mu3}\right| \left|U_{\mu2}\right| \left|U_{e3}\right| \left|U_{e2}\right|
\sin\left(\varphi^{e}_{23} - \varphi^{\mu}_{23} \right)\,.
\end{equation}
In what follows we will use the three Majorana-type phases $\varphi^{e}_{23}\,$,  $\varphi^{\mu}_{23}\,$ and  $\varphi^{\tau}_{23}\,$.

\begin{table}[t]
\begin{center}
\caption{\label{tab:data} Neutrino oscillation parameter summary from 
ref.~\cite{Tortola:2012te}. For $\Delta m^2_{31}$, $\sin^2\theta_{23}$, $\sin^2\theta_{13}$, and
$\delta$ the upper (lower) row corresponds to normal (inverted)
neutrino mass hierarchy.} \small
\begin{tabular}{|c|c|c|}
\hline
parameter & best fit & 3$\sigma$ range \\
\hline && \\[1mm]
$\Delta m^2_{21} \left[10^{-5}\text{ eV}\right]$ & 7.62  &
$7.12-8.20$ \\[3mm]
$|\Delta m^2_{31}|\: \left[10^{-3}\text{ eV}\right]$   &
\begin{tabular}{c}
2.55\\
2.43
\end{tabular}    &
\begin{tabular}{c}
$2.31-2.74$\\
$2.21-2.64$
\end{tabular} \\[6mm]
$\sin^2\theta_{12}$ & 0.320 & $0.27-0.37$\\[3mm]
$\sin^2\theta_{23}$    &
\begin{tabular}{c}
0.613 (0.427)\\
0.600
\end{tabular}
&
\begin{tabular}{c}
$0.36-0.68$\\
$0.37-0.67$
\end{tabular}\\[5mm]
$\sin^2\theta_{13}$
&
\begin{tabular}{c}
0.0246\\
0.0250
\end{tabular}
&
$0.017-0.033$ \\[5mm]

$\delta$
&
\begin{tabular}{r}
$0.80\pi$\\
$-0.03\pi$
\end{tabular}
&
$0-2\pi$ \\[5mm]
\hline
\end{tabular}
\end{center}
\end{table}

In our analysis we have calculated $O_e$ numerically using the charged
lepton masses given at $M_Z$ scale in the $\overline{MS}$ scheme at
1-loop~\cite{Fusaoka:1998vc,Xing:2007fb} as
\begin{subequations}
\label{eq:lepmasses}
\begin{align}
  m_e&=0.486661305\pm{0.000000056}\text{ MeV}\,,\\
  m_{\mu}&=102.728989\pm{0.000013}\text{ MeV}\,,\\
  m_{\tau}&=1746.28\pm{0.16}\text{ MeV}\,.
 \end{align}
\end{subequations}
Concerning the neutrino sector we have used the most recent three
neutrino data from the global fit of neutrino oscillations in
ref.~\cite{Tortola:2012te}. The best fit values and 3$\sigma$
ranges for the neutrino parameters are presented in table~\ref{tab:data}.

As in ref.~\cite{EmmanuelCosta:2011jq}, here we have varied all the
experimental charged lepton masses and neutrino mass differences
within their allowed range given in eq.~\eqref{eq:lepmasses} and
Table~\ref{tab:data}, respectively.  The mass of the lightest neutrino
($m_1$ in NH or $m_3$ in IH) was scanned for different magnitudes
below 1 eV. To reconstruct the PMNS matrix, we have also scanned the free
parameters $A_e\,$, $B_e\,$, $D_{\nu}\,$ and the phases
$\kappa_1,\,\kappa_2,\,\varphi$, defined in eq.~\eqref{eq:lep}. All
the remaining parameters are calculated in terms of the former
ones. The restriction in this scan was to accept only the input values
which correspond to a reconstructed PMNS matrix~$U$ that naturally
leads to the mixing angles $\theta_{12}\,,\theta_{23}$ and
$\theta_{13}$ within their experimental bounds presented in
Table~\ref{tab:data}.

\begin{figure}[h]
  \begin{center}
    \includegraphics[width=7cm,clip]{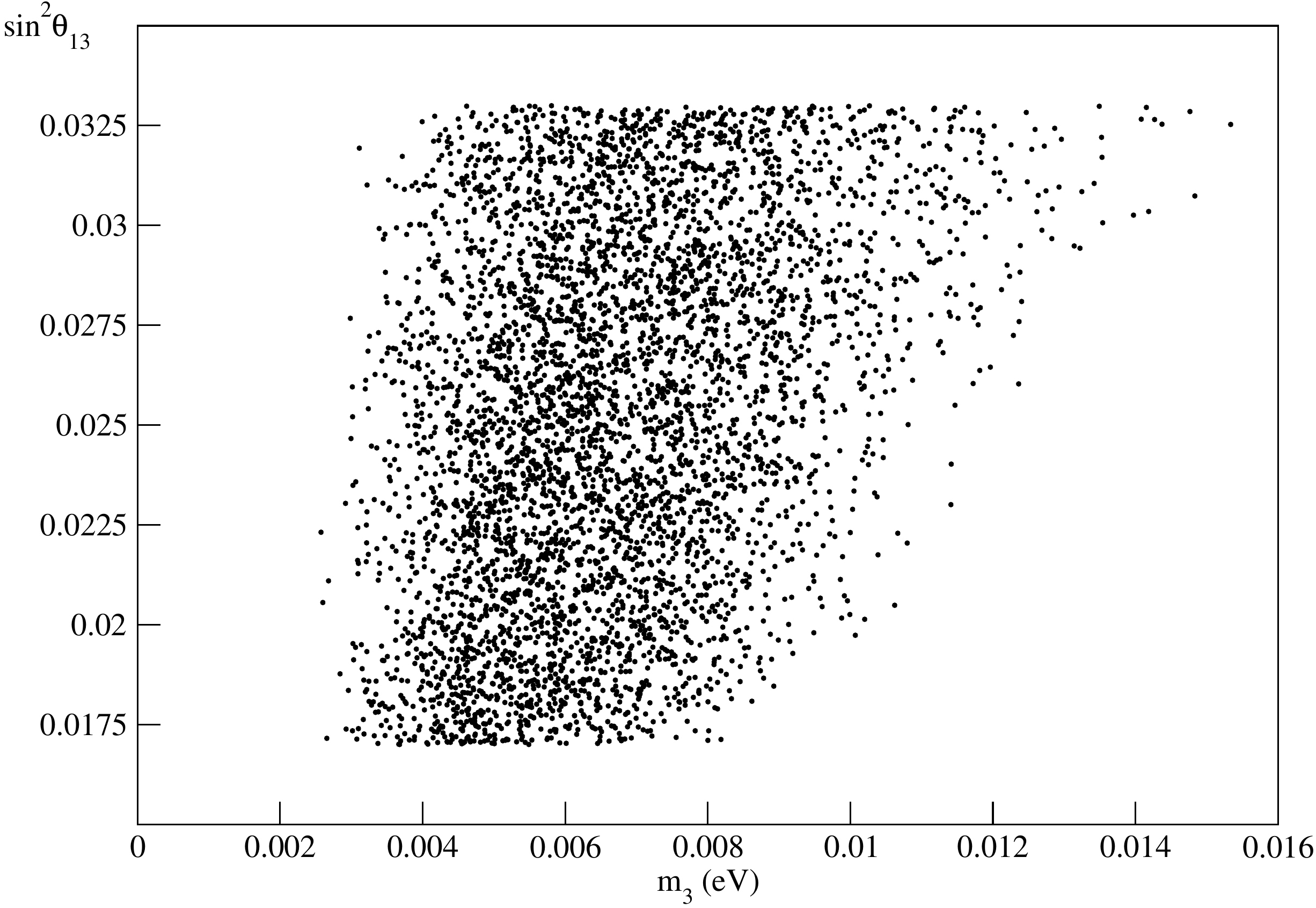}
    \qquad
    \includegraphics[width=7cm,clip]{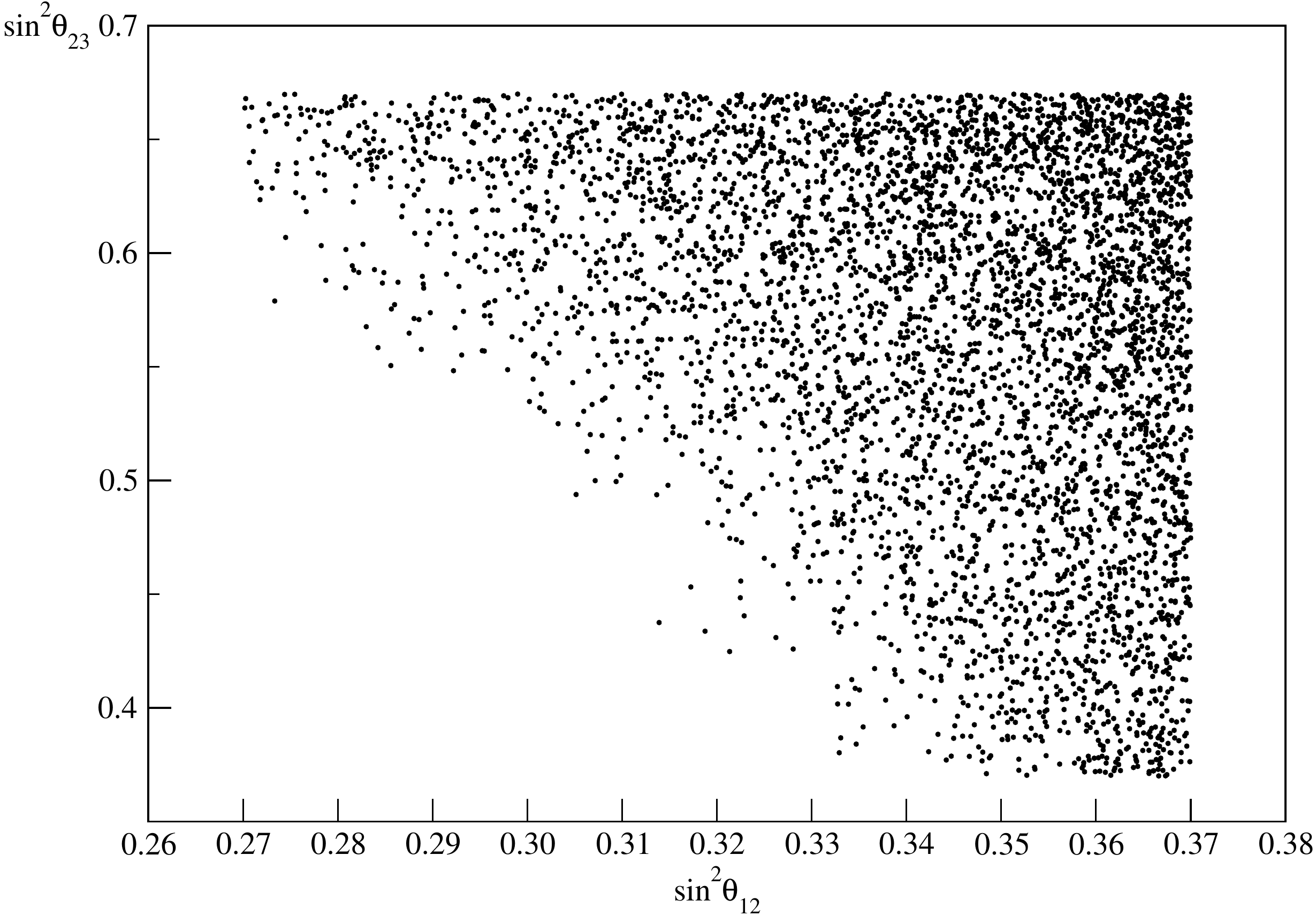}
  \end{center}
 \caption{\label{fig4} Plot of $\sin^2 \theta_{13}$ as a function of 
 m$_3$ (left panel) and $\sin^2 \theta_{23}$ as a function of 
 $\sin^2 \theta_{12}$ (right) in the case of texture $m^{A_{\scriptscriptstyle(12)}}_\nu$ 
 and inverted hierarchy.} 
\end{figure}

From our scan we have found that the mass matrix $A$ in
eq.~\eqref{eq:three} is compatible with a neutrino mass spectrum with
normal hierarchy while the texture $A_{\scriptscriptstyle(12)}$ is
compatible with inverted hierarchy. The allowed ranges for the
lightest neutrino masses are $m_1=[0.353,20.884]\times10^{-3}$~eV for
NH and $m_3=[2.575,15.335]\times10^{-3}$~eV for IH. The presence of a
massless neutrino as well as a quasi-degenerate neutrino mass spectrum
are excluded in both cases. For the texture
$A$ we have found no significant correlations between $\sin^2
\theta_{13}$ and $\sin^2 \theta_{23}$ as a function of $m_1$, while in
the case of texture $A_{\scriptscriptstyle(12)}$ some correlations are
found, as shown in figure~\ref{fig4}. In fact, this correlation is behind the
narrower $m_3$ allowed range for IH in comparison with the allowed $m_1$ range for NH.
We have also verified that textures $A$ and $A_{\scriptscriptstyle(12)}$ are not compatible with
inverted and normal hierarchies, respectively, even when the new
limits on $\sin^2 \theta_{13}$ are considered.

\begin{figure}[t]
  \begin{center}
    \includegraphics[width=14cm,clip]{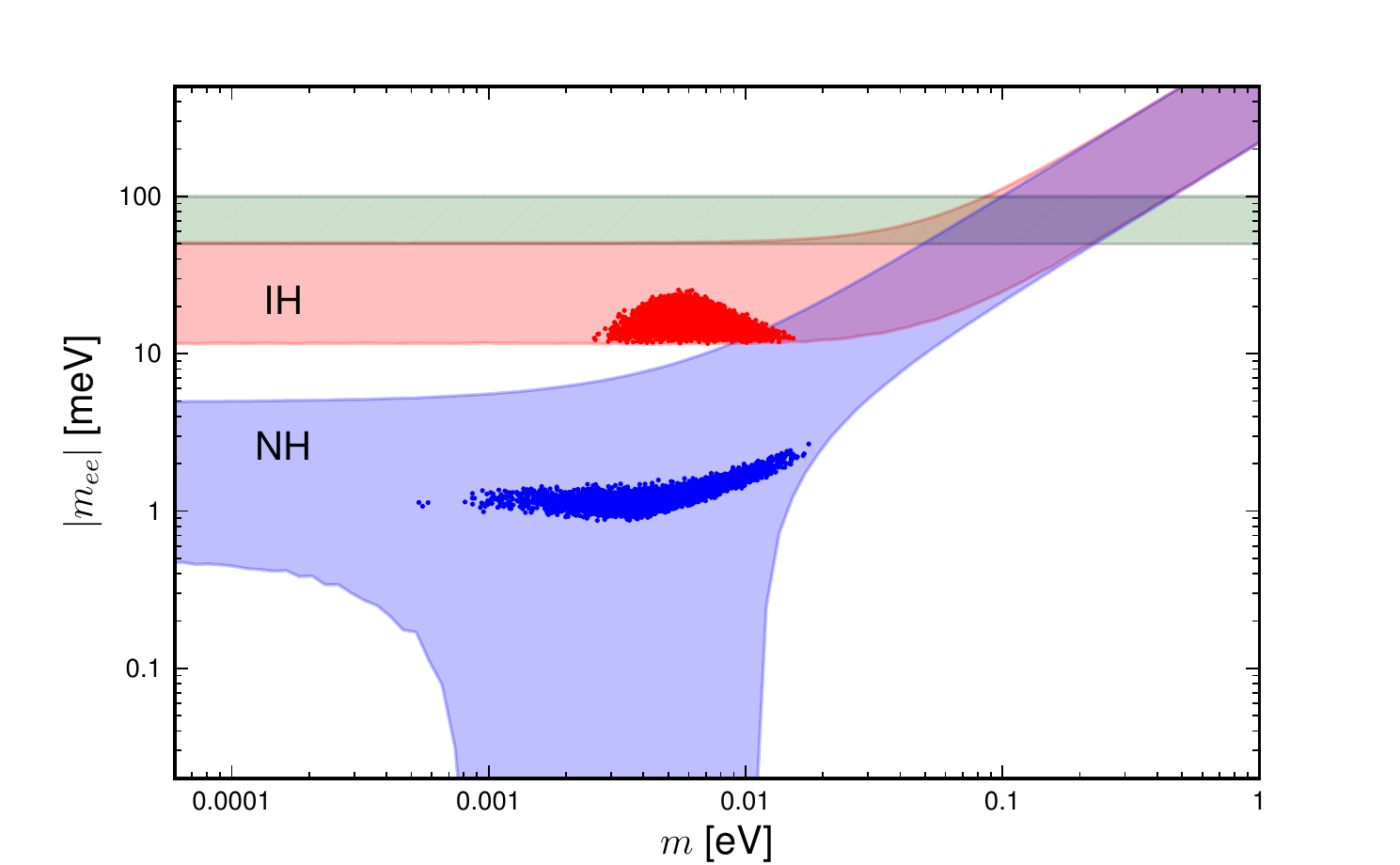}
  \end{center}
  \caption{\label{fig3} Effective Majorana neutrino mass $m_{ee}$ as a
    function of the lightest neutrino mass $m$ for normal and inverted
    neutrino hierarchy, as indicated. The upper band shows the
    experimental sensitivity to be achieved in the next years.}  
\end{figure}

In figure ~\ref{fig3} we plot the effective Majorana neutrino mass
characterizing the neutrinoless double beta decay amplitude $m_{ee}$
with respect to the lightest neutrino mass, $m_1$ in the case of
$m^{A}_\nu$ and normal hierarchy or $m_3$ in the case of
$m^{A_{\scriptscriptstyle(12)}}_\nu$ and inverted hierarchy.  The
shadowed bands correspond to the generic predictions for $m_{ee}$
according to the experimental neutrino data at 3$\sigma$, without any
further assumption concerning the origin of neutrino masses. If we now
restrict our calculations to the adjoint $\mathsf{SU}(5)\times
\mathsf{Z}_4$ model presented in this article, the allowed regions are
reduced to the darker pointed regions.  The horizontal lines in figure
~\ref{fig3} correspond to the $m_{ee}$ sensitivity that will be
reached by the next generation of neutrinoless double beta decay
experiments (see for instance ref.~\cite{Giuliani:2012zu}). There we
see that, even if a part of the inverse hierarchy band will be
experimentally covered in the next years, a sensitivity of around
10-30 meV will be needed in order to probe the effective Majorana mass
predicted by our model.  Accessing to the predicted region for normal
hierarchy will be even tougher, since a sensitivity of the order of 1
meV would be required, far away from the more optimistic scenarios.
 
\section{Conclusions}
\label{sec:conclusions}

In this work we have studied the adjoint-$\mathsf{SU(5)}\times
\mathsf{Z}_4$ model. The flavour symmetry imposed in the Lagrangian
has the purpose to force the quark mass matrices $M_u,\,M_d$ to have
the NNI form after the spontaneous electroweak symmetry
breaking~\cite{Branco:2010tx,EmmanuelCosta:2011jq}. Due to the fermion
content of the adjoint-$\mathsf{SU(5)}$, the charged lepton mass
matrix $M_e$ has automatically NNI form. In this model the light
neutrinos get their masses through type-I, type-III and one-loop
radiative seesaw mechanisms, implemented, respectively, via a singlet,
a triplet and an octet from the adjoint fermionic fields. The
$\mathsf{SU(5)}\times \mathsf{Z}_4$ symmetry does not impose any
constraint on the adjoint fermionic fields.

We have shown that the model proposed is in agreement with the current
experimental data. Neutrino mixings and mass splittings as well as the
masses of the charged leptons have been used to constrain the possible
textures of the effective light neutrino mass matrix. We have
demonstrated that at least three copies of the $\mathsf{24}$ are
needed in order to fully implement the $\mathsf{Z}_4$ flavour symmetry
and simultaneously account for the experimental neutrino data. As
shown in table~\ref{tab:results} only two zero-textures persist: $A$
and $A_{\scriptscriptstyle(12)}$, which are compatible with normal and
inverted hierarchies, respectively.

One of the main phenomenological implications of the model studied is
the prediction of a hierarchical neutrino mass spectrum not compatible
with a massless neutrino. This result is particularly important since
the neutrino mass spectrum predicted can be used to prove or disprove
the model in the near future. At present our results are in agreement
with the constraints coming from neutrinoless double beta
decay~\cite{Giuliani:2012zu} and tritium $\beta$ decay
searches~\cite{Otten:2008zz} as well as with the cosmological bound on
the sum of light neutrino masses~\cite{Ade:2013xsa}. However, a
positive signal of neutrinoless double beta decay in the next years as
well as a cosmological measurement of the sum of neutrino masses of
the order of 0.1 eV would certainly rule out this type of
model. Therefore, future experimental improvements in the neutrino
physics will be decisive for testing the viability of the
$\mathsf{SU(5)}\times\mathsf{Z}_4$ model.

\acknowledgments The work of C.~S. is supported by Funda\c{c}\~ao para
a Ci\^encia e a Tecnologia (FCT, Portugal) under the contract
SFRH/BD/61623/2009. The work of D.E.C. was supported by Associa\c
c\~ao do Instituto Superior T\'ecnico para a Investiga\c c\~ao e
Desenvolvimento (IST-ID).  The work of D.E.C. and C.S. was also
supported by Portuguese national funds through FCT - Funda\c c\~ao
para a Ci\^encia e Tecnologia, project PEst-OE/FIS/UI0777/2011, Marie
Curie RTN MRTN-CT-2006-035505 and by Funda\c c\~ao para a Ci\^encia e a
Tecnologia (FCT, Portugal) through the projects CERN/FP/83503/2008,
PTDC/FIS/ 098188/2008, CERN/FP/116328/2010 and CFTP-FCTUNIT 777.
M.T.~acknowledges financial support from CSIC under the JAE-Doc
programme, co-funded by the European Social Fund and from the Spanish
MINECO under grants FPA2011-22975 and MULTIDARK CSD2009-00064
(Consolider-Ingenio 2010 Programme) as well as from the Generalitat
Valenciana grant Prometeo/2009/091.

\appendix
\section{Matter and Higgs representations}
\label{reps}

\subsubsection*{Fermionic representations}
 The fermionic fields in the model decompose in terms of the SM gauge quantum numbers as:
\begin{equation}
  \begin{aligned}
    \mathsf{\mathsf{5}^{\ast}}&=d^c\,\oplus\,L\,,\\
    \mathsf{10}&=Q\,\oplus\,u^c\,\oplus\,e^c\,,\\
    \mathsf{24}&=\rho_8\,\oplus\,\rho_3\,\oplus\,\rho_{(3,2)}\,\oplus\,\rho_{(\overline{3},2)}\,\oplus\,
    \rho_0\,.
  \end{aligned}
\end{equation}
The fermionic representations $\mathsf{5}^\ast$ and $\mathsf{10}$ can be written as
\begin{equation}
\label{eq:vev7}
 \mathsf{5}^{\ast}_\alpha=(d^c)_\alpha\,, \quad \mathsf{5}^{\ast}_i=\varepsilon_{ij}l^j\,,
\end{equation}
and 
\begin{equation}
  10^{\alpha\beta}=\frac{1}{\sqrt{2}}\varepsilon^{\alpha \beta \gamma}(u^c)_\gamma,\quad 10^{\alpha i}=-\frac{1}{\sqrt{2}}q^{\alpha i},\quad 10^{ij}=\frac{1}{\sqrt{2}}\varepsilon^{ij}e^c\,,
  \label{eq:ten}
\end{equation}
where $\alpha,\,\beta,\,\gamma=1,2,3$ and $i,j=4,5$.
The fermionic field $\rho_3$, triplet of $\mathsf{SU(2)}$, 
belonging to the adjoint representation can be written as,
\begin{equation}
  \rho_3=\frac{1}{2}\begin{pmatrix} 
    {\rho^{}_3}^{0} & \sqrt{2}{\rho^{}_3}^{+}\\[2mm]
    \sqrt{2}{\rho^{}_3}^{-} & -{\rho^{}_3}^0
  \end{pmatrix}\,,
\end{equation}
where
\begin{equation}
  {\rho^{}_3}^{\pm}=\frac{\rho^1_3\mp i\rho^2_3}{\sqrt{2}}\,,\qquad 
  {\rho^{}_3}^{0}=\rho^3_3\,.
\end{equation}


\subsubsection*{Higgs representations}

The Higgs content of the model decomposes as
\begin{equation}
  \label{eq:higgs}
  \begin{aligned}
    \mathsf{5}_H&=T_1\,\oplus\,H_1\,,\\
    \mathsf{24}_H&=\Sigma_8\,\oplus\,\Sigma_3\,\oplus\,\Sigma_{(3,2)}\,\oplus\,\Sigma_{(\overline{3},2)}
    \,\oplus\,\Sigma_0\,,\\
    \mathsf{45}_H&=S_{(8,2)_{\frac{3}{10}}}\,\oplus\,S_{(\bar{6},1)_{-\frac{1}{5}}}\,\oplus\,S_{(3,3)_{-\frac{1}{5}}}\,\oplus\,S_{(\bar{3},2)_{-\frac{7}{10}}} 
\,\oplus\,S_{(\bar{3},1)_{\frac{4}{5}}}\,\oplus\,T_2\,\oplus\,H_2\,,
  \end{aligned}
\end{equation}
where we have included for completeness the hypercharged properly
normalised. The $H_1$ and $H_2$ are the usual Higgs doublets and $T_1$
and $T_2$ are the colour triplets. The $\mathsf{45}$ Higgs representation, which the explicit decomposition 
is given in ref.~\cite{Kannike:2011fx}, obeys to the following relations,
\begin{equation}
\label{eq:vev2}
  \mathsf{45}^{ij}_k=-\mathsf{45}^{ji}_k \qquad \text{and} \qquad \sum_{j=1}^{5}\mathsf{45}^{ij}_j=0\,,
\end{equation}

The different contributions for the beta coefficients $b_i$ of each
extra particle besides the 2HDM content are given in
table~\ref{tab:bi}.

\begin{table}[h]
\begin{center}
  \caption{\label{tab:bi} Summary of the $b_i$ constants for relevant
    particles in the model.}
\begin{tabular}{ccccccccc}
\\\hline\hline
 & 2HDM & $\rho_3$ & $\rho_8$ & $\rho_{(3,2)}$ & $T_{1,2}$ & $\Sigma_3$ & $\Sigma_8$ & $S_{(8,2)}$  \\
\hline\\[-2mm]
$b_1$ &  21/5 & 0 & 0 & $\frac{5}{3}$ & $\frac{1}{15}$  & 0 & 0 & $\frac{4}{5}$ \\[2mm]
$b_2$ & -3 & $\frac{4}{3}$ & 0 & 1 & 0 & $\frac{2}{3}$ & 0 & $\frac{4}{3}$  \\[2mm]
$b_3$ & -7 & 0 & 2 & $\frac{2}{3}$ & $\frac{1}{6}$ & 0 & 1 & 2  \\[2mm]
\hline\hline
\end{tabular}
\end{center}
\end{table}

\section{The Potential}
\label{sec:pot}

In this section we give explicitly the terms of the Higgs
potential. Notice that index $H$ on the Higgs fields is dropped
in the following expressions. The potential $V$  is divided into six parts as
follows:
\begin{equation}
\begin{split}
  V\left(\mathsf{5}, \mathsf{24}, \mathsf{45}\right) =& V_1\left(\mathsf{5}\right) \,+\,
  V_2\left(\mathsf{24}\right) \,+\, V_3\left(\mathsf{45}\right) \,+\, 
V_4\left(\mathsf{24},\mathsf{45}\right)\\[2mm] \,+\,&
V_5\left(\mathsf{5},\mathsf{24}\right)\,+\,
V_6\left(\mathsf{5},\mathsf{45}\right)\,,
\end{split}
\end{equation}
where each parcell are given by:
\begin{subequations}
\begin{equation}
  V_1\left(\mathsf{5}\right)\,=\,-\frac{\mu_{\mathsf{5}}^2}{2}\,\mathsf{5}^{\alpha}\,\mathsf{5}^{\ast}_{\alpha}
  \,+\, \frac{\lambda_1}{4}\,\left(\mathsf{5}^{\alpha}\,\mathsf{5}^{\ast}_{\alpha}\right)^2\,,
\end{equation}
\begin{equation}
  \begin{split}
    V_2\left(\mathsf{24}\right)\,=&%
    -\frac{\mu^2_{\mathsf{24}}}{2}\,
    \mathsf{24}^\alpha_{\beta}\,\mathsf{24}^\beta_{\alpha} +
    \frac{\lambda_2}{2}\left(\mathsf{24}^\alpha_{\beta}\,
      \mathsf{24}^\beta_{\alpha}\right)^2
    \,+\, \frac{a_{1}}{3}\,\mathsf{24}^\alpha_{\beta}\, \mathsf{24}^\beta_{\gamma}\, \mathsf{24}^\gamma_{\alpha}\\[2mm]
    & \,+\, \frac{\lambda_{3}}{2}\,\mathsf{24}^\alpha_{\beta}\,
    \mathsf{24}^\beta_{\gamma}\, \mathsf{24}^\gamma_{\delta}\,
    \mathsf{24}^\delta_{\alpha}\,,
  \end{split}
\end{equation}
\begin{equation}
  \begin{split}
    \label{eq:pot45}
    V_3\left(\mathsf{45}\right)=&%
    -\frac{\mu^2_{\mathsf{45}}}{2}\,{\mathsf{45}}^{\alpha
      \beta}_\gamma {\mathsf{45}^{\ast}}^\gamma_{\alpha \beta} 
    \,+\, \lambda_4 \left(\mathsf{45}^{\alpha \beta}_\gamma
      {\mathsf{45}^{\ast}}^\gamma_{\alpha \beta}\right)^2 \\[2mm]
    \,+\, & \lambda_5 \mathsf{45}^{\alpha \beta}_\gamma
    {\mathsf{45}^{\ast}}^\delta_{\alpha \beta} \mathsf{45}^{\kappa
      \lambda}_\delta
    {\mathsf{45}^{\ast}}^\gamma_{\kappa \lambda} 
    \,+\, \lambda_6 \mathsf{45}^{\alpha \beta}_\gamma
    {\mathsf{45}^{\ast}}^\delta_{\alpha \beta}\mathsf{45}^{\kappa
      \gamma}_{\lambda} {\mathsf{45}^{\ast}}^{\lambda}_{\kappa \delta}\\[2mm]
    \,+\, & \lambda_7 \mathsf{45}^{\alpha \delta}_\beta
    {\mathsf{45}^{\ast}}^\beta_{\alpha \gamma} \mathsf{45}^{\kappa
      \gamma}_{\lambda} {\mathsf{45}^{\ast}}^{\lambda}_{\kappa
      \delta}
    \,+\, \lambda_8 \mathsf{45}^{\alpha \gamma}_\delta
    {\mathsf{45}^{\ast}}^\beta_{\gamma \epsilon} \mathsf{45}^{\kappa
      \delta}_\alpha {\mathsf{45}^{\ast}}^\epsilon_{\kappa\beta}\\[2mm]
    \,+\,& \lambda_9 \mathsf{45}^{\alpha \gamma}_\delta
    {\mathsf{45}^{\ast}}^{\beta}_{\gamma \epsilon}
    \mathsf{45}^{\kappa \epsilon}_\alpha
    {\mathsf{45}^{\ast}}^{\delta}_{\kappa \beta}
    \,+\, \lambda_{10} \mathsf{45}^{\alpha \gamma}_\delta
    {\mathsf{45}^{\ast}}^{\beta}_{\gamma \epsilon}
    \mathsf{45}^{\kappa \delta}_\beta
    {\mathsf{45}^{\ast}}^{\epsilon}_{\kappa \alpha} \\[2mm]
    \,+\, & \lambda_{11} \mathsf{45}^{\alpha \gamma}_\delta
    {\mathsf{45}^{\ast}}^{\beta}_{\gamma \epsilon}
    \mathsf{45}^{\kappa \epsilon}_\beta
    {\mathsf{45}^{\ast}}^\delta_{\kappa \alpha}\,,
  \end{split}
\end{equation}
\begin{equation}
  \begin{split}
    V_4\left(\mathsf{24},\mathsf{45}\right)\,  = &%
    a_2\, \mathsf{45}^{\alpha \beta}_\gamma
    \mathsf{24}^\gamma_{\delta}
    {\mathsf{45}^{\ast}}^{\delta}_{\alpha \beta} 
    \,+\,
    \lambda_{12}\, \mathsf{45}^{\alpha \beta}_\gamma\,
    {\mathsf{45}^{\ast}}^\gamma_{\alpha \beta}\,
    \mathsf{24}^\delta_{\epsilon}
    \mathsf{24}^\epsilon_{\delta}\\[2mm]
    \,+\, & \lambda_{13}\, \mathsf{45}^{\alpha \beta}_\gamma
    \mathsf{24}^\delta_{\alpha} \mathsf{24}^\epsilon_{\beta}
    {\mathsf{45}^{\ast}}^\gamma_{\delta \epsilon}
    \,+\, \lambda_{14}\, \mathsf{45}^{\alpha \beta}_\gamma
    \mathsf{24}^\gamma_{\beta} \mathsf{24}^\delta_{\epsilon}
    {\mathsf{45}^{\ast}}^\epsilon_{\alpha \delta} \\[2mm]
    \,+\, & \lambda_{15}\, \mathsf{45}^{\alpha \beta}_\gamma
    \mathsf{24}^\gamma_{\epsilon} \mathsf{24}^\delta_{\beta}
    {\mathsf{45}^{\ast}}^\epsilon_{\alpha \delta} 
    \,+\, \lambda_{16} \mathsf{45}^{\alpha \beta}_\gamma
    \mathsf{24}^{\kappa}_{\alpha} \mathsf{24}^\lambda_{\kappa}
    {\mathsf{45}^{\ast}}^\gamma_{\lambda
      \beta} \\[2mm]
    \,+\, & \lambda_{17}\, \mathsf{45}^{\alpha \beta}_\gamma
    \mathsf{24}^\gamma_{\kappa} \mathsf{24}^{\kappa}_{\lambda}
    {\mathsf{45}^{\ast}}^\lambda_{\alpha \beta}\,,
  \end{split}
\end{equation}
\begin{equation}
    V_5\left(\mathsf{5},\mathsf{24}\right)\,=\,
    a_3\, \mathsf{5}^{\ast}_\alpha \mathsf{24}^\alpha_{\beta}
    \mathsf{5}^\beta 
    \,+\, \lambda_{18}\, \mathsf{5}^{\ast}_\alpha \mathsf{5}^\alpha
    \mathsf{24}^\beta_{\gamma} \mathsf{24}^\gamma_{\beta}
    \,+\, \lambda_{19} \, \mathsf{5}^{\ast}_\alpha
    \mathsf{24}^\alpha_{\beta} \mathsf{24}^\beta_{\gamma} \mathsf{5}^\gamma\,,
\end{equation}
and
\begin{equation}
    \label{eq:pot45e5}
  V_6\left(\mathsf{5}, \mathsf{45}\right)\,=\,
  \lambda_{20}\,\mathsf{45}^{\alpha \beta}_\gamma {\mathsf{45}^{\ast}}^\gamma_{\alpha \beta} \mathsf{5}^{\ast}_\delta \mathsf{5}^\delta
  \,+\, \lambda_{21}\, \mathsf{45}^{\alpha \beta}_\delta \mathsf{5}^{\ast}_\gamma {\mathsf{45}^{\ast}}^\gamma_{\alpha \beta} \mathsf{5}^\delta
  \,+\, \lambda_{22}\, \mathsf{45}^{\alpha \beta}_{\gamma} {\mathsf{45}^{\ast}}^{\gamma}_{\alpha \delta} \mathsf{5}^{\ast}_\beta
  \mathsf{5}^\delta\,.
\end{equation}
\end{subequations}

\addcontentsline{toc}{section}{References}
\bibliographystyle{JHEP}
\bibliography{refs}

\end{document}